\documentclass[sigplan,screen]{acmart}

\AtBeginDocument{%
  }

\renewcommand\footnotetextcopyrightpermission[1]

\usepackage{algorithm}
\usepackage{algpseudocode}

\usepackage{tikz}
\usepackage{amsmath}
\usepackage{subfigure}
\usepackage{multirow}

\newcommand{\method}[0]{FlexInfer}

\begin{document}

\title[FlexInfer: Breaking Memory Constraint for On-Device LLM Inference]{FlexInfer: Breaking Memory Constraint via Flexible and Efficient Offloading for On-Device LLM Inference}


\author{Hongchao Du}
\authornote{These authors contributed equally to this work.}
\affiliation{%
  \institution{City University of Hong Kong}
  \city{Hong Kong}
  \country{China}
}

\author{Shangyu Wu}
\authornotemark[1]
\affiliation{%
  \institution{City University of Hong Kong}
  \city{Hong Kong}
  \country{China}
}

\author{Arina Kharlamova}
\affiliation{%
  \institution{MBZUAI}
  \city{Abu Dhabi}
  \country{United Arab Emirates}
}


\author{Nan Guan}
\affiliation{%
 \institution{City University of Hong Kong}
 \city{Hong Kong}
 \country{China}
}

\author{Chun Jason Xue}
\affiliation{%
  \institution{MBZUAI}
  \city{Abu Dhabi}
  \country{United Arab Emirates}
}
\begin{abstract}
Large Language Models (LLMs) face challenges for on-device inference due to high memory demands. 
Traditional methods to reduce memory usage often compromise performance and lack adaptability. 
We propose~\method, an optimized offloading framework for on-device inference, addressing these issues with techniques like asynchronous prefetching, balanced memory locking, and flexible tensor preservation. These strategies enhance memory efficiency and mitigate I/O bottlenecks, ensuring high performance within user-specified resource constraints. Experiments demonstrate that FlexInfer significantly improves throughput under limited resources, achieving up to 12.5 times better performance than existing methods and facilitating the deployment of large models on resource-constrained devices.

\end{abstract}





\maketitle
\section{Introduction}
The success of Large Language Models (LLMs) has revolutionized numerous fields, enabling breakthroughs in natural language understanding, generation, and decision-making tasks~\cite{openai2024openaio1card,23arxiv-llama,deepseekai2025}. 
However, existing LLMs are typically deployed on powerful cloud-based infrastructures, which may introduce many significant issues, such as privacy concerns~\cite{powerinfer, PrivateGPT}, and lack of customization~\cite{lyu2024llmrec}.
Deploying LLMs on edge devices is gaining a growing interest~\cite{alizadeh2024llmflash, xu2023llmcad, xue2024powerinfer2, powerinfer}, particularly in scenarios where sensitive data handling, model customization, and independent operation are crucial.

Performing on-device inference still faces significant challenges due to the substantial memory demands of LLMs~\cite{kaplan2020scalinglaws}, which often exceed the capacities of local devices~\cite{wang-etal-2024-compression,mao2024compressibility}. 
To reduce resource demands, existing methods propose strategies such as distilling smaller models~\cite{hsieh2023distilling,zhang2024tinyllama}, applying model compression~\cite{MLSYS2024_42a452cb,MLSYS2024_5edb57c0,xu2024onebit}, and pruning models~\cite{alizadeh2024llmflash, powerinfer,xue2024powerinfer2, xia2023flashllm, pmlr-v202-liu23am, 24arxiv-raee,wu2025evoprobustllminference}. 
Although these approaches can improve models' memory efficiency, they inevitably impact the generality performance and still suffer in extreme resource-constrained scenarios~\cite{chen2024rolesmallmodelsllm,Huang_2024, he-etal-2024-chess}. 
Furthermore, these methods lack the flexibility to vary memory budgets or deployment constraints, requiring adjusting the hyper-parameters, such as quantization or sparsity levels, offering limited choices, and imposing overhead on adjustments.

To address memory limitations, several works leverage external storage to supplement limited device memory~\cite{alizadeh2024llmflash, powerinfer, xue2024powerinfer2}. 
A typical way is to offload model parameters to storage devices and fetch them on demand~\cite{llama.cpp, 23icml-flexgen, xue2024powerinfer2}.
However, inefficiently performing I/O operations between memory and storage would slow the inference~\cite{xue2024powerinfer2,Mobile_IO}.
Moreover, existing offloading methods typically do not support flexible memory usage, so they also have limited adaptability to varying resource constraints. 
To provide flexibility for various resource-constrained environments, this paper proposes~\textbf{\method}, a memory-efficient LLM inference offloading framework. 
\method~first introduces \textbf{asynchronous prefetching} to alleviate I/O overheads and parallelize I/O operations and computations, then proposes \textbf{balanced memory locking} to uniformly retain model parameters to make full use of available memory. \method~also presents \textbf{flexible tensor preservation} to determine what model parameters should be offloaded and retained based on user-specified resource budgets.
Those techniques perform precise memory management and enhance IO efficiency.
We conducted extensive experiments to show that the proposed \method~can achieve 10.6-12.5 times inference speedup compared to existing offloading techniques across various memory-limited scenarios.


In summary, this paper makes the following contributions:
\begin{itemize}
\item We propose \method, a novel framework that optimizes offloading-based on-device inference for LLMs through asynchronous prefetching, balanced memory locking, and flexible tensor preservation.
\item We develop precise memory management strategies to minimize I/O bottlenecks and maximize memory efficiency, enabling the deployment of large models on resource-limited environments.
\item Extensive experiments demonstrate that FlexInfer significantly outperforms existing methods with high throughput under varying user-specified budgets.
\end{itemize}

\section{Background and Motivations}

\subsection{LLM Inference}

Existing LLMs basically adopt the transformer-based architecture~\cite{17nips-trans}, which consists of multiple transformer blocks. Each transformer block contains a self-attention module and a feedforward module. 
Given a token sequence $X=[x_1, x_2, \ldots, x_n]$, where each $x_i$ is a $d$-dimensional vector, the self-attention module computes three internal states, i.e., $Q=XW_Q^T$, $K=XW_K^T$, and $V=XW^T_V$, then the outputs of the attention modules can be computed as, 
\begin{equation}
    Attention(Q, K, V)=softmax\left(\frac{QK^T}{\sqrt{d_k}}\right)V,
\end{equation}
where $d_k$ is the dimension of the keys. 
For the feedforward module, it computes the output with two linear transformations and a nonlinear activation function.
\begin{equation}
    FFN(h)=ACT(hW_{up}+b_{up})W_{down}+b_{down},
\end{equation}
where $h$ is the outputs of the attention modules, $ACT(\cdot)$ is the activation function such as SwiGLU in Llama-series models~\cite{23arxiv-llama, 23arxiv-llama-2, 24arxiv-llama-3}. $W_Q, W_K, W_V, W_{up}, b_{up}, W_{down}, b_{down}$ are learnable parameters, and all of these parameters are required when generating each token.


\subsection{Deployment on Memory-Constrained Devices}

Mobile devices and edge computing platforms are crucial to enable real-time, low-latency interactions with LLMs. 
However, these devices are typically constrained in terms of memory and processing power compared to cloud-based systems or high-performance servers. 
The massive number of parameters in modern LLMs often exceeds the memory capabilities of these devices. 
For example, state-of-the-art models such as Llama series models~\cite{24arxiv-llama-3,23arxiv-llama,23arxiv-llama-2} have billions of parameters, resulting in memory footprints that can easily surpass the available memory of most mobile devices.


Several strategies have been proposed to reduce memory and computational demands, e.g., adopting smaller models~\cite{hsieh2023distilling,zhang2024tinyllama}, model compression~\cite{MLSYS2024_42a452cb,MLSYS2024_5edb57c0,xu2024onebit}, and model pruning~\cite{alizadeh2024llmflash, powerinfer,xue2024powerinfer2, xia2023flashllm, pmlr-v202-liu23am,wu2025evoprobustllminference}. While these approaches reduce resource requirements, they share two critical limitations: \textbf{Lack of Flexibility}: These methods cannot directly adapt to varying memory budgets or constraints. The required memory size becomes fixed once parameters such as model size, quantization levels, or sparsity thresholds are determined. \textbf{Limited Scalability}: Despite these optimizations, large-scale models that exceed available memory remain unsupported. 
To support models exceeding available memory, offloading-based methods reduce memory usage by moving parts of data stored in the memory to external or slower memory. However, this will bring non-negligible IO overhead, and careful design is required to reduce performance loss.

\begin{table}[t]
\caption{Inference throughput (tokens/second) on memory-constrained scenarios. The size of `Llama2-70B' is about 36.2 GB, and the full-memory throughput is 31.14 tokens/s.}
\label{tab:motivation}
\begin{center}
\begin{tabular}{ccccccccc}
\toprule
\textbf{Ava Mem}&\textbf{5} & \textbf{10} & \textbf{15} & \textbf{20} & \textbf{25} & \textbf{30} & \textbf{35}\\
\midrule
Llama2-70B&0.51 & 0.49 & 0.49 & 0.46 & 0.50 & 1.41 & 2.06 \\
\bottomrule
\end{tabular}
\end{center}
\end{table}

\subsection{Motivations}

To show the impact of deploying memory offloading in LLM inference under memory-constrained scenarios, we conducted preliminary experiments with the state-of-the-art inference engine llama.cpp~\cite{llama.cpp} and used cgroup~\cite{cgroup} to limit the available memory.
The default serving method in llama.cpp is mmap~\cite{mmap}, which loads the corresponding data from the storage with page faults if it is not in the memory when accessed.
Table~\ref{tab:motivation} presents the inference throughput of the 4-bit quantized llama-2-70b chat model~\cite{23arxiv-llama-2} under various memory-constrained scenarios. 
The results indicate a substantial decrease in inference performance when memory is not sufficient.
This is because almost every access to the model weight triggers IO operations.
Providing more memory improves the inference performance to some extent, but it is still far from the performance of full memory. In summary, the offload-based on-device inference must address the following challenges: high-cost IO operations, inability to utilize available memory fully, and lack of flexibility and scalability for different memory sizes. 
\section{\method}


This section introduces the design of \method, an offloading-based framework for efficient on-device inference of LLMs under resource constraints. We first present the system architecture (\S\ref{overview}), followed by the asynchronous prefetching mechanism that optimizes the overlap of I/O and computation (\S\ref{async_prefetch}). We then detail the balanced memory locking strategy for efficient memory management (\S\ref{memory_retention}) and the flexible tensor preservation technique for intelligent parameter selection (\S\ref{tensor_preservation}). Finally, section~\ref{implementation} discusses implementation considerations.

\begin{figure}[t]
  \centering
  \includegraphics[]{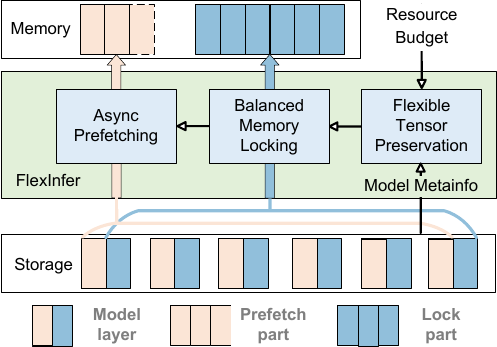}
  \caption{~\method~architecture.}
  \label{fig:architecture}
  \Description{\method~architecture.}
\end{figure}

\subsection{Architecture Overview}
\label{overview}
As illustrated in Figure~\ref{fig:architecture}, \method~consists of three key components, which operate collaboratively during inference: the flexible tensor preservation optimizer first determines the parameter preservation plan based on the available budget and model metadata information, and the memory locking manager then divides the model into two parts: one is loaded and fixed in memory, while the other is loaded on demand through prefetching, handled by the asynchronous prefetch module. The detailed design and implementation of each component are presented in the following sections.




\subsection{Asynchronous Prefetching: Reducing I/O Overhead}


When sufficient memory is available, all model parameters
can reside in memory, eliminating the need for storage I/O operations. 
If memory constraints prevent full model loading, offloading-based methods leverage storage devices as an extension of memory, dynamically loading required parameters. 
In this scenario, inference performance is significantly influenced by I/O overhead,
the model throughput (tokens/s) with synchronous offloading can be expressed as\footnote{This paper focuses on CPU inference since resource-constrained devices usually don't have powerful GPUs.}:
\begin{equation}
T_{sync} = \frac{1}{Per\_token\_CPU\_latency + \frac{IO\_size}{IO\_bandwidth}}
\end{equation}
One common optimization is to parallelize I/O and computation operations, leading to an improved theoretical performance model:
\begin{equation}
T_{async} = \frac{1}{\max(Per\_token\_CPU\_latency, \frac{IO\_size}{IO\_bandwidth})}\label{equation:async}
\end{equation}
This formulation reveals that optimal performance in offloading-based methods depends on two key factors: maximizing the parallelization between I/O and computation threads and fully utilizing available storage bandwidth.

\label{async_prefetch}
\begin{figure}[t]
  \includegraphics[]{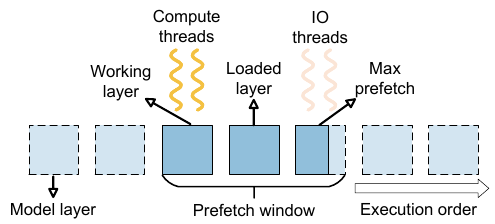}
  \caption{Asynchronous prefetching.}
  \label{fig:async_prefetch}
  \Description{Asynchronous prefetching.}
\end{figure}

\method~employs a tensor-based multi-threaded prefetching strategy to achieve efficient parallelization and high bandwidth. We maintain input and output embedding layers in memory, focusing our strategy on handling identical-sized decoding layers. The computation threads process a layer only after the I/O threads load their parameters, with synchronization managed through atomic operations on shared variables. Multiple IO threads or computing threads will collaborate to process a particular layer and move to the next layer together, with each I/O thread responsible for loading a single tensor (e.g., $W_Q, W_K, W_V$.). This multi-threaded IO operation at the tensor-level granularity helps avoid inefficient small-size data transfers, optimizing bandwidth utilization.

A key observation in large model inference is that each parameter is accessed precisely once during token generation, eliminating any potential for parameter locality optimization. Our offloading strategy leverages this characteristic by immediately releasing the memory after parameter usage. The total memory footprint is thus determined by the size of the prefetch window, as shown in Figure~\ref{fig:async_prefetch}. 
Consequently, our offloading method achieves a memory reduction ratio of approximately $\frac{k}{n}$ compared to the original model size, where $n$ represents the number of model layers and $k$ is the prefetch window size.

\subsection{Balanced Memory Locking: Maximizing Memory Efficiency}
\label{memory_retention}
The offloading-based method effectively reduces memory usage but is constrained by IO overhead, which limits performance. Moreover, increasing the available memory does not improve performance, as prefetching alone cannot reduce the required IO during each inference cycle. To address this challenge, \method~introduces an adaptive memory-locking strategy. This strategy leverages excess memory to retain specific parameters in memory, reducing the \textit{IO\_size} in formula~\ref{equation:async} and bringing better performance. However, a critical aspect of this approach is selecting the appropriate parameters to retain in memory for optimal results.

\begin{figure}[t]
  \centering
  \includegraphics[]{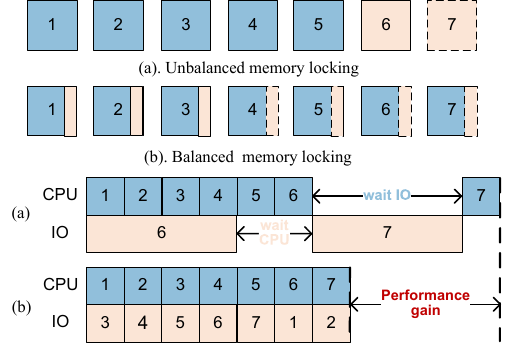}
  \caption{Balanced  Memory Locking.}
  \label{fig:adaptive_memory}
  \Description{Balanced Memory Locking.}
\end{figure}

A naive approach might involve retaining several layers of the model in memory, thereby directly removing the IO needed for those layers. However, such an uneven memory-locking strategy introduces variability in processing speed across layers, causing the computation and IO threads to wait on each other. For example, as illustrated in Figure~\ref{fig:adaptive_memory}(a), a layer-based memory-locking method retains the first five layers entirely in memory while using the remaining memory to store one of the last two layers. This imbalance leads to IO thread delays until the compute thread releases memory and subsequent compute thread delays while waiting for IO to complete. This hinders the system's ability to achieve complete parallelism between IO and computation, ultimately affecting overall performance.

To address this issue, the balanced memory-locking strategy divides each layer into two parts: one fixed in memory and the other dynamically prefetched, as shown in Figure~\ref{fig:adaptive_memory}(b). By distributing memory usage uniformly, the IO workload for each layer remains stable throughout the inference process, enabling consistent and efficient parallelization of IO and computation. By maintaining a balanced locking approach, \method~achieves significantly better performance, minimizing unnecessary delays and maximizing resource utilization.

\subsection{Flexible Tensor Preservation: Heuristic Parameter Management}
\label{tensor_preservation}


When the tensor sizes within each model layer are the same size, balanced memory locking can evenly distribute available memory across all layers. However, varying tensor sizes in LLMs affect performance depending on which parameters are kept in memory. 
The transformer architecture provides a consistent tensor structure across LLMs. Most parameters are associated with attention tensors ($W_Q, W_K, W_V$) and FFN tensors ($W_{up}, b_{up}, W_{down}, b_{down}$), typically with an approximate 1:3 size ratio between one attention tensor and one FFN tensor. Attention and MLP tensors have their own advantages and disadvantages under different available memory sizes. For example, when available memory is small, prioritizing attention parameters can save as many tensors as possible in memory, reduce the number of IO operations, and implement larger-size IO through FFN tensors. When memory is considerable, saving all FFN tensors can minimize memory fragmentation and the difference in residual size between layers, keeping each layer's IO overhead uniform.
\begin{algorithm}[t]
\caption{Flexible Tensor Preservation Algorithm.}
\label{alg1}
\begin{algorithmic}[1]
\Statex \textbf{Input:} Attention tensor size $size_{atte}$, FFN tensor size $size_{FFN}$, Layer number $N$, Memory budget $size_{mem}$, 
\Statex \textbf{Output:} Tensor preservation plan $P$
\If{$size_{mem} >= size_{FFN} * N * 3 + size_{attn} * N * 2$}
    \State Set all FFN tensors for all layers
\Else
    \If{$size_{mem} >= size_{FFN} * N * 2$}
        \State Set two FFN tensor for all layers
    \Else
        \If{$size_{mem} >= size_{FFN} * N$}
            \State Set one FFN tensor for all layers
        \EndIf
    \EndIf
\EndIf
\State Set as much as possible attention tensors one by one 
\State return $P$
\end{algorithmic}
\end{algorithm}

\begin{figure*}[t]
  \includegraphics[]{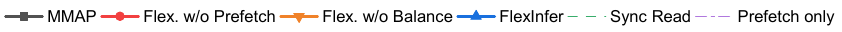}\\
  \centering
  \subfigure[Llama2-7B (3.8GB, 12.72 tokens/s)]{\includegraphics[]{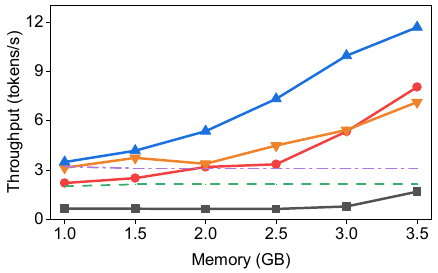}\label{fig:throughput-7B}}
  \subfigure[Llama2-13B (7.3GB, 6.74 tokens/s)]{\includegraphics[]{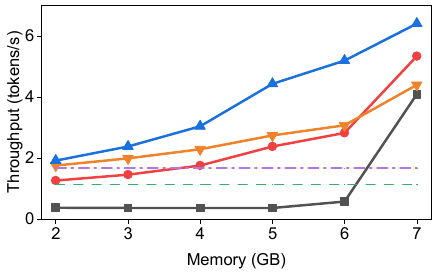}\label{fig:throughput-13B}}
  \\
  \subfigure[Codellama-34B (17.9GB, 2.6 tokens/s)]{\includegraphics[]{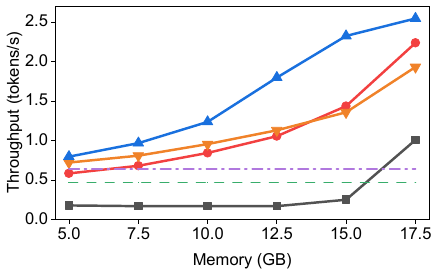}\label{fig:throughput-34B}}
  \subfigure[Llama2-70B (36.4GB, 1.3 tokens/s)]{\includegraphics[]{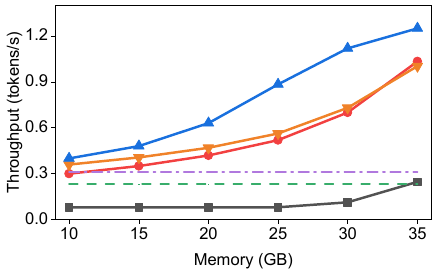}\label{fig:throughput-70B}}
  \caption{Evaluation result.}
  \label{fig:throughput-8t}
  \Description{Evaluation result.}
\end{figure*}
Leveraging this predictable tensors structure of LLMs, we developed a heuristic algorithm, outlined in Algorithm~\ref{alg1}. The core idea is to select parameters to retain in memory based on the available memory size, which is shown as follows:
\begin{itemize}
    \item When memory is sufficient: If the available memory can accommodate all FFN parameters and half attention tensors, FFN tensors are prioritized and fully retained.
    \item When memory is limited: Attention tensors are prioritized if memory is insufficient to hold one FFN parameter for all layers.
    \item Intermediate cases: When memory falls between these two extremes, FFN parameters are selected incrementally until the remaining memory cannot hold one FFN tensor for all layers. At that point, as many attention parameters as possible are retained.\footnote{For models that apply GQA, we prefer smaller $W_k,W_v$ compared to $W_q,W_o$.}
\end{itemize}
This heuristic ensures that the remaining parameters across layers and any unused memory fragments do not exceed the size of a single attention tensor. When the remaining memory cannot hold an extra tensor for all layers, we prioritize the attention tensor to reduce the differences across layers and prioritize the large-size IO of FFN. This design balances memory utilization and IO efficiency, optimizing the inference process with a simple strategy.

\subsection{Implementation}
\label{implementation}

The \method~framework was implemented with 828 lines of C/C++ code, extending llama.cpp~\cite{llama.cpp} to incorporate asynchronous prefetching, balanced memory locking, and flexible tensor preservation. Additional parameters were introduced to control the available memory and configure the number of threads. 
The default size of the prefetch window is set to 3, ensuring efficient memory management and minimal latency during inference. 
We use direct IO to bypass the page cache for IO threads.

\section{Evaluation}
\subsection{Experimental Setup}
To show the results under different available memory conditions, we tested the performance of \method~on an Ubuntu server with 512GB memory and AMD 7995WX CPU.
We used cgroup~\cite{cgroup} and teskset~\cite{taskset} to limit the available memory and CPU cores to simulate resource-constrained devices. 
In addition to the mmap baseline (\textit{MMAP}), we tested \method~without prefetching (\textit{Flex. w/o Prefetch}) and \method~without balanced memory locking (\textit{Flex. w/o Balance}) under different available memory conditions to show the effect of prefetching and balanced locking. \textit{Flex. w/o Prefetch} loads all parameters synchronously into memory before use and \textit{Flex. w/o Balance} locks the model parameters by layer order. At the same time, to show the impact of adaptive memory locking, we also tested the results of reading parameters synchronously separately (\textit{Sync Read}) and prefetching separately (\textit{Prefetch only}).

\subsection{Inference Throughput}

The decoding performance of Llama2 series models~\cite{23arxiv-llama-2}, including llama2-7B, llama2-13B, Codellama-34B, and llama2-70B, are shown in Figure~\ref{fig:throughput-8t}. The total memory size required to run each model and the performance when sufficient memory is available are shown in the title of each subfigure. Experimental results show that although mmap can run with very little memory, it can only achieve very limited inference performance, with only 0.08-0.67 tokens/s for different models. When the available memory increases, the performance slightly improves. This is because mmap causes all parameters to be loaded into memory through inefficient synchronous IO, and the parameters loaded into memory are swapped out of memory before the next use, resulting in very limited scalability brought by more memory. In contrast, \method~effectively improves the inference performance under different memory conditions, achieving performance improvements of 5.2-12.5x, 5-11.8x, 4.2-10.6x, and 5-11x than mmap under models of different sizes. 

\subsection{Ablation Studies}

By replacing mmap with multiple threads large reads (\textit{Sync Read}), \method~can improve performance by 2.6-3x when memory is limited, proving that IO is the main bottleneck for large model inference under memory-constrained conditions. By introducing prefetching, \method~further improves performance by 34.8-59.4\% by parallelizing computation and IO. As available memory increases, the performance improvement achieved by prefetching can reach up to 69.9-118.8\%. Balanced locking has a limited improvement over unbalanced memory locking when memory is low, ranging from 9.2-11.1\% at minimum memory settings. This is because only a small number of parameters are locked in memory, so the difference between different strategies is limited. With more memory, balanced locking can improve by up to 56.8-83.3\%.

\begin{figure}[t]
  \centering
  \subfigure[Llama2-7B]{\includegraphics[]{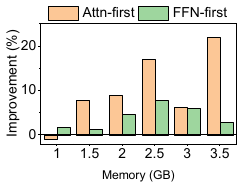}\label{fig:ablation-7B}}
  \subfigure[Llama2-13B]{\includegraphics[]{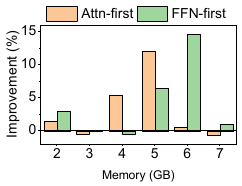}\label{fig:ablation-13B}}
  \caption{Ablation study for flexible tensor preservation.}
  \label{fig:ablation}
  \Description{Ablation study for flexible tensor preservation.}
  \vspace{-0.5cm}
\end{figure}

For the ablation experiment of the parameter preservation algorithm, we compared two simple strategies, \textit{Attn-first} and \textit{FFN-first}, which prioritize the attention parameters and FFN parameters to be retained in memory. Figure 5 shows the performance improvement of our method over the simple strategy at different memory sizes. We only provide results for the 7B and 13B models since the 34B and 70B models employ GQA~\cite{gqa}, which makes our strategy produce the same optimal results as \textit{Attn-first} in most cases. Experimental results show that \method~can achieve up to 21.9\% and 7.8\% performance improvement on 7B and 13B models, respectively, compared with Attn-first, and 12\% and 14.6\% performance improvement compared with FFN-first.
\section{Related Work}
\subsection{LLM Inference Serving Engines}
With the rapid development of LLM, many model-serving systems~\cite{FasterTransformer,ocra,pope2022efficientlyscalingtransformerinference,Pets,wang-etal-2021-lightseq} and model-serving optimization~\cite{deepspeed,flexgen,zero-offload,Superneurons,MLSYS2020_0b816ae8,pmlr-v162-patil22b,FlasAttention,pagedattention,accelerate,DejaVu} have been proposed. However, most of them are optimized for server environments with powerful GPUs and batch-based workloads. Among them, Hugging Face Accelerate~\cite{accelerate} and DeepSpeed Zero~\cite{deepspeed} support offloading model parameters to CPU memory or SSD. However, they still rely on GPU and are unsuitable for running on edge devices. FlexGen~\cite{flexgen} studies swapping model parameters between GPU, CPU, and SSD to alleviate memory requirements but still targets batch-based workloads and requires at least one GPU. Inspired by virtual memory technology, vLLM~\cite{pagedattention} proposes paged attention to alleviate the memory waste and inefficiency problems of LLM at runtime. Still, it is also aimed at batch-based online service scenarios. In contrast, FlexInfer can be applied to any edge devices and optimized for on-device local inference.

\subsection{Offloading for On-device Inference}
For inference on edge devices, several offloading-based methods that do not rely on GPUs have been proposed~\cite{alizadeh2024llmflash, powerinfer,xue2024powerinfer2,llama.cpp}. Llama.cpp~\cite{llama.cpp} is an LLM inference engine implemented in C/C++. It supports running in multiple environments and uses mmap to achieve offloading on SSD. However, its performance is unsatisfactory under memory constraints and lacks scalability. Sparsity-based selective model loadings such as LLMFlash~\cite{alizadeh2024llmflash} and PowerInfer~\cite{powerinfer,xue2024powerinfer2} significantly improve inference performance by reducing the IO size at the algorithm level. However, their performance improvement depends on sparsity and may impact model capabilities. In contrast, FlexInfer can be directly applied to any transform-based model while maintaining full capabilities. Other inference optimization methods, such as model compression~\cite{MLSYS2024_42a452cb,MLSYS2024_5edb57c0,xu2024onebit}, model quantization~\cite{alizadeh2024llmflash, powerinfer,xue2024powerinfer2, xia2023flashllm, pmlr-v202-liu23am}, speculative decoding~\cite{LLMCad,speculative_decoding}, etc., are orthogonal to the method proposed in this paper.

\section{Conclusion}

Large Language Models (LLMs) pose significant challenges for local inference in resource-constrained environments. This paper addresses these challenges by introducing \method, a novel framework that combines asynchronous prefetching, balanced memory locking, and flexible tensor preservation to optimize offloading-based LLM inference. \method~minimizes I/O overhead and maximizes memory efficiency under diverse user-specified resource budgets. Extensive experiments demonstrate its superiority over existing methods, making it a practical solution for deploying LLMs locally. This work paves the way for more flexible, efficient, and accessible LLM applications in privacy-sensitive and local scenarios.



\bibliographystyle{ACM-Reference-Format}
\bibliography{refer}


\begin{thebibliography}{51}


\ifx \showCODEN    \undefined \def \showCODEN     #1{\unskip}     \fi
\ifx \showDOI      \undefined \def \showDOI       #1{#1}\fi
\ifx \showISBNx    \undefined \def \showISBNx     #1{\unskip}     \fi
\ifx \showISBNxiii \undefined \def \showISBNxiii  #1{\unskip}     \fi
\ifx \showISSN     \undefined \def \showISSN      #1{\unskip}     \fi
\ifx \showLCCN     \undefined \def \showLCCN      #1{\unskip}     \fi
\ifx \shownote     \undefined \def \shownote      #1{#1}          \fi
\ifx \showarticletitle \undefined \def \showarticletitle #1{#1}   \fi
\ifx \showURL      \undefined \def \showURL       {\relax}        \fi
\providecommand\bibfield[2]{#2}
\providecommand\bibinfo[2]{#2}
\providecommand\natexlab[1]{#1}
\providecommand\showeprint[2][]{arXiv:#2}

\bibitem[Ainslie et~al\mbox{.}(2023)]%
        {gqa}
\bibfield{author}{\bibinfo{person}{Joshua Ainslie}, \bibinfo{person}{James Lee-Thorp}, \bibinfo{person}{Michiel de Jong}, \bibinfo{person}{Yury Zemlyanskiy}, \bibinfo{person}{Federico Lebrón}, {and} \bibinfo{person}{Sumit Sanghai}.} \bibinfo{year}{2023}\natexlab{}.
\newblock \bibinfo{title}{GQA: Training Generalized Multi-Query Transformer Models from Multi-Head Checkpoints}.
\newblock
\newblock
\showeprint[arxiv]{2305.13245}~[cs.CL]
\urldef\tempurl%
\url{https://arxiv.org/abs/2305.13245}
\showURL{%
\tempurl}


\bibitem[Alizadeh et~al\mbox{.}(2024)]%
        {alizadeh2024llmflash}
\bibfield{author}{\bibinfo{person}{Keivan Alizadeh}, \bibinfo{person}{Iman Mirzadeh}, \bibinfo{person}{Dmitry Belenko}, \bibinfo{person}{Karen Khatamifard}, \bibinfo{person}{Minsik Cho}, \bibinfo{person}{Carlo C~Del Mundo}, \bibinfo{person}{Mohammad Rastegari}, {and} \bibinfo{person}{Mehrdad Farajtabar}.} \bibinfo{year}{2024}\natexlab{}.
\newblock \bibinfo{title}{LLM in a flash: Efficient Large Language Model Inference with Limited Memory}.
\newblock
\newblock
\showeprint[arxiv]{2312.11514}~[cs.CL]
\urldef\tempurl%
\url{https://arxiv.org/abs/2312.11514}
\showURL{%
\tempurl}


\bibitem[Aminabadi et~al\mbox{.}(2022)]%
        {deepspeed}
\bibfield{author}{\bibinfo{person}{Reza~Yazdani Aminabadi}, \bibinfo{person}{Samyam Rajbhandari}, \bibinfo{person}{Ammar~Ahmad Awan}, \bibinfo{person}{Cheng Li}, \bibinfo{person}{Du Li}, \bibinfo{person}{Elton Zheng}, \bibinfo{person}{Olatunji Ruwase}, \bibinfo{person}{Shaden Smith}, \bibinfo{person}{Minjia Zhang}, \bibinfo{person}{Jeff Rasley}, {and} \bibinfo{person}{Yuxiong He}.} \bibinfo{year}{2022}\natexlab{}.
\newblock \showarticletitle{DeepSpeed- Inference: Enabling Efficient Inference of Transformer Models at Unprecedented Scale}. In \bibinfo{booktitle}{\emph{SC22: International Conference for High Performance Computing, Networking, Storage and Analysis}}. \bibinfo{pages}{1--15}.
\newblock
\urldef\tempurl%
\url{https://doi.org/10.1109/SC41404.2022.00051}
\showDOI{\tempurl}


\bibitem[Chen and Varoquaux(2024)]%
        {chen2024rolesmallmodelsllm}
\bibfield{author}{\bibinfo{person}{Lihu Chen} {and} \bibinfo{person}{Gaël Varoquaux}.} \bibinfo{year}{2024}\natexlab{}.
\newblock \bibinfo{title}{What is the Role of Small Models in the LLM Era: A Survey}.
\newblock
\newblock
\showeprint[arxiv]{2409.06857}~[cs.CL]
\urldef\tempurl%
\url{https://arxiv.org/abs/2409.06857}
\showURL{%
\tempurl}


\bibitem[Dao et~al\mbox{.}(2022)]%
        {FlasAttention}
\bibfield{author}{\bibinfo{person}{Tri Dao}, \bibinfo{person}{Dan Fu}, \bibinfo{person}{Stefano Ermon}, \bibinfo{person}{Atri Rudra}, {and} \bibinfo{person}{Christopher R\'{e}}.} \bibinfo{year}{2022}\natexlab{}.
\newblock \showarticletitle{FlashAttention: Fast and Memory-Efficient Exact Attention with IO-Awareness}. In \bibinfo{booktitle}{\emph{Advances in Neural Information Processing Systems}}, \bibfield{editor}{\bibinfo{person}{S.~Koyejo}, \bibinfo{person}{S.~Mohamed}, \bibinfo{person}{A.~Agarwal}, \bibinfo{person}{D.~Belgrave}, \bibinfo{person}{K.~Cho}, {and} \bibinfo{person}{A.~Oh}} (Eds.), Vol.~\bibinfo{volume}{35}. \bibinfo{publisher}{Curran Associates, Inc.}, \bibinfo{pages}{16344--16359}.
\newblock
\urldef\tempurl%
\url{https://proceedings.neurips.cc/paper_files/paper/2022/file/67d57c32e20fd0a7a302cb81d36e40d5-Paper-Conference.pdf}
\showURL{%
\tempurl}


\bibitem[DeepSeek-AI et~al\mbox{.}(2025)]%
        {deepseekai2025}
\bibfield{author}{\bibinfo{person}{DeepSeek-AI}, \bibinfo{person}{Daya Guo}, \bibinfo{person}{Dejian Yang}, \bibinfo{person}{Haowei Zhang}, \bibinfo{person}{Junxiao Song}, \bibinfo{person}{Ruoyu Zhang}, \bibinfo{person}{Runxin Xu}, \bibinfo{person}{Qihao Zhu}, \bibinfo{person}{Shirong Ma}, \bibinfo{person}{Peiyi Wang}, \bibinfo{person}{Xiao Bi}, \bibinfo{person}{Xiaokang Zhang}, \bibinfo{person}{Xingkai Yu}, \bibinfo{person}{Yu Wu}, \bibinfo{person}{Z.~F. Wu}, \bibinfo{person}{Zhibin Gou}, \bibinfo{person}{Zhihong Shao}, \bibinfo{person}{Zhuoshu Li}, \bibinfo{person}{Ziyi Gao}, \bibinfo{person}{Aixin Liu}, \bibinfo{person}{Bing Xue}, \bibinfo{person}{Bingxuan Wang}, \bibinfo{person}{Bochao Wu}, \bibinfo{person}{Bei Feng}, \bibinfo{person}{Chengda Lu}, \bibinfo{person}{Chenggang Zhao}, \bibinfo{person}{Chengqi Deng}, \bibinfo{person}{Chenyu Zhang}, \bibinfo{person}{Chong Ruan}, \bibinfo{person}{Damai Dai}, \bibinfo{person}{Deli Chen}, \bibinfo{person}{Dongjie Ji}, \bibinfo{person}{Erhang Li},
  \bibinfo{person}{Fangyun Lin}, \bibinfo{person}{Fucong Dai}, \bibinfo{person}{Fuli Luo}, \bibinfo{person}{Guangbo Hao}, \bibinfo{person}{Guanting Chen}, \bibinfo{person}{Guowei Li}, \bibinfo{person}{H. Zhang}, \bibinfo{person}{Han Bao}, \bibinfo{person}{Hanwei Xu}, \bibinfo{person}{Haocheng Wang}, \bibinfo{person}{Honghui Ding}, \bibinfo{person}{Huajian Xin}, \bibinfo{person}{Huazuo Gao}, \bibinfo{person}{Hui Qu}, \bibinfo{person}{Hui Li}, \bibinfo{person}{Jianzhong Guo}, \bibinfo{person}{Jiashi Li}, \bibinfo{person}{Jiawei Wang}, \bibinfo{person}{Jingchang Chen}, \bibinfo{person}{Jingyang Yuan}, \bibinfo{person}{Junjie Qiu}, \bibinfo{person}{Junlong Li}, \bibinfo{person}{J.~L. Cai}, \bibinfo{person}{Jiaqi Ni}, \bibinfo{person}{Jian Liang}, \bibinfo{person}{Jin Chen}, \bibinfo{person}{Kai Dong}, \bibinfo{person}{Kai Hu}, \bibinfo{person}{Kaige Gao}, \bibinfo{person}{Kang Guan}, \bibinfo{person}{Kexin Huang}, \bibinfo{person}{Kuai Yu}, \bibinfo{person}{Lean Wang}, \bibinfo{person}{Lecong Zhang},
  \bibinfo{person}{Liang Zhao}, \bibinfo{person}{Litong Wang}, \bibinfo{person}{Liyue Zhang}, \bibinfo{person}{Lei Xu}, \bibinfo{person}{Leyi Xia}, \bibinfo{person}{Mingchuan Zhang}, \bibinfo{person}{Minghua Zhang}, \bibinfo{person}{Minghui Tang}, \bibinfo{person}{Meng Li}, \bibinfo{person}{Miaojun Wang}, \bibinfo{person}{Mingming Li}, \bibinfo{person}{Ning Tian}, \bibinfo{person}{Panpan Huang}, \bibinfo{person}{Peng Zhang}, \bibinfo{person}{Qiancheng Wang}, \bibinfo{person}{Qinyu Chen}, \bibinfo{person}{Qiushi Du}, \bibinfo{person}{Ruiqi Ge}, \bibinfo{person}{Ruisong Zhang}, \bibinfo{person}{Ruizhe Pan}, \bibinfo{person}{Runji Wang}, \bibinfo{person}{R.~J. Chen}, \bibinfo{person}{R.~L. Jin}, \bibinfo{person}{Ruyi Chen}, \bibinfo{person}{Shanghao Lu}, \bibinfo{person}{Shangyan Zhou}, \bibinfo{person}{Shanhuang Chen}, \bibinfo{person}{Shengfeng Ye}, \bibinfo{person}{Shiyu Wang}, \bibinfo{person}{Shuiping Yu}, \bibinfo{person}{Shunfeng Zhou}, \bibinfo{person}{Shuting Pan}, \bibinfo{person}{S.~S. Li},
  \bibinfo{person}{Shuang Zhou}, \bibinfo{person}{Shaoqing Wu}, \bibinfo{person}{Shengfeng Ye}, \bibinfo{person}{Tao Yun}, \bibinfo{person}{Tian Pei}, \bibinfo{person}{Tianyu Sun}, \bibinfo{person}{T. Wang}, \bibinfo{person}{Wangding Zeng}, \bibinfo{person}{Wanjia Zhao}, \bibinfo{person}{Wen Liu}, \bibinfo{person}{Wenfeng Liang}, \bibinfo{person}{Wenjun Gao}, \bibinfo{person}{Wenqin Yu}, \bibinfo{person}{Wentao Zhang}, \bibinfo{person}{W.~L. Xiao}, \bibinfo{person}{Wei An}, \bibinfo{person}{Xiaodong Liu}, \bibinfo{person}{Xiaohan Wang}, \bibinfo{person}{Xiaokang Chen}, \bibinfo{person}{Xiaotao Nie}, \bibinfo{person}{Xin Cheng}, \bibinfo{person}{Xin Liu}, \bibinfo{person}{Xin Xie}, \bibinfo{person}{Xingchao Liu}, \bibinfo{person}{Xinyu Yang}, \bibinfo{person}{Xinyuan Li}, \bibinfo{person}{Xuecheng Su}, \bibinfo{person}{Xuheng Lin}, \bibinfo{person}{X.~Q. Li}, \bibinfo{person}{Xiangyue Jin}, \bibinfo{person}{Xiaojin Shen}, \bibinfo{person}{Xiaosha Chen}, \bibinfo{person}{Xiaowen Sun}, \bibinfo{person}{Xiaoxiang
  Wang}, \bibinfo{person}{Xinnan Song}, \bibinfo{person}{Xinyi Zhou}, \bibinfo{person}{Xianzu Wang}, \bibinfo{person}{Xinxia Shan}, \bibinfo{person}{Y.~K. Li}, \bibinfo{person}{Y.~Q. Wang}, \bibinfo{person}{Y.~X. Wei}, \bibinfo{person}{Yang Zhang}, \bibinfo{person}{Yanhong Xu}, \bibinfo{person}{Yao Li}, \bibinfo{person}{Yao Zhao}, \bibinfo{person}{Yaofeng Sun}, \bibinfo{person}{Yaohui Wang}, \bibinfo{person}{Yi Yu}, \bibinfo{person}{Yichao Zhang}, \bibinfo{person}{Yifan Shi}, \bibinfo{person}{Yiliang Xiong}, \bibinfo{person}{Ying He}, \bibinfo{person}{Yishi Piao}, \bibinfo{person}{Yisong Wang}, \bibinfo{person}{Yixuan Tan}, \bibinfo{person}{Yiyang Ma}, \bibinfo{person}{Yiyuan Liu}, \bibinfo{person}{Yongqiang Guo}, \bibinfo{person}{Yuan Ou}, \bibinfo{person}{Yuduan Wang}, \bibinfo{person}{Yue Gong}, \bibinfo{person}{Yuheng Zou}, \bibinfo{person}{Yujia He}, \bibinfo{person}{Yunfan Xiong}, \bibinfo{person}{Yuxiang Luo}, \bibinfo{person}{Yuxiang You}, \bibinfo{person}{Yuxuan Liu}, \bibinfo{person}{Yuyang Zhou},
  \bibinfo{person}{Y.~X. Zhu}, \bibinfo{person}{Yanhong Xu}, \bibinfo{person}{Yanping Huang}, \bibinfo{person}{Yaohui Li}, \bibinfo{person}{Yi Zheng}, \bibinfo{person}{Yuchen Zhu}, \bibinfo{person}{Yunxian Ma}, \bibinfo{person}{Ying Tang}, \bibinfo{person}{Yukun Zha}, \bibinfo{person}{Yuting Yan}, \bibinfo{person}{Z.~Z. Ren}, \bibinfo{person}{Zehui Ren}, \bibinfo{person}{Zhangli Sha}, \bibinfo{person}{Zhe Fu}, \bibinfo{person}{Zhean Xu}, \bibinfo{person}{Zhenda Xie}, \bibinfo{person}{Zhengyan Zhang}, \bibinfo{person}{Zhewen Hao}, \bibinfo{person}{Zhicheng Ma}, \bibinfo{person}{Zhigang Yan}, \bibinfo{person}{Zhiyu Wu}, \bibinfo{person}{Zihui Gu}, \bibinfo{person}{Zijia Zhu}, \bibinfo{person}{Zijun Liu}, \bibinfo{person}{Zilin Li}, \bibinfo{person}{Ziwei Xie}, \bibinfo{person}{Ziyang Song}, \bibinfo{person}{Zizheng Pan}, \bibinfo{person}{Zhen Huang}, \bibinfo{person}{Zhipeng Xu}, \bibinfo{person}{Zhongyu Zhang}, {and} \bibinfo{person}{Zhen Zhang}.} \bibinfo{year}{2025}\natexlab{}.
\newblock \bibinfo{title}{DeepSeek-R1: Incentivizing Reasoning Capability in LLMs via Reinforcement Learning}.
\newblock
\newblock
\showeprint[arxiv]{2501.12948}~[cs.CL]
\urldef\tempurl%
\url{https://arxiv.org/abs/2501.12948}
\showURL{%
\tempurl}


\bibitem[Dubey et~al\mbox{.}(2024)]%
        {24arxiv-llama-3}
\bibfield{author}{\bibinfo{person}{Abhimanyu Dubey}, \bibinfo{person}{Abhinav Jauhri}, \bibinfo{person}{Abhinav Pandey}, \bibinfo{person}{Abhishek Kadian}, \bibinfo{person}{Ahmad Al{-}Dahle}, \bibinfo{person}{Aiesha Letman}, \bibinfo{person}{Akhil Mathur}, \bibinfo{person}{Alan Schelten}, \bibinfo{person}{Amy Yang}, \bibinfo{person}{Angela Fan}, \bibinfo{person}{Anirudh Goyal}, \bibinfo{person}{Anthony Hartshorn}, \bibinfo{person}{Aobo Yang}, \bibinfo{person}{Archi Mitra}, \bibinfo{person}{Archie Sravankumar}, \bibinfo{person}{Artem Korenev}, \bibinfo{person}{Arthur Hinsvark}, \bibinfo{person}{Arun Rao}, \bibinfo{person}{Aston Zhang}, \bibinfo{person}{Aur{\'{e}}lien Rodriguez}, \bibinfo{person}{Austen Gregerson}, \bibinfo{person}{Ava Spataru}, \bibinfo{person}{Baptiste Rozi{\`{e}}re}, \bibinfo{person}{Bethany Biron}, \bibinfo{person}{Binh Tang}, \bibinfo{person}{Bobbie Chern}, \bibinfo{person}{Charlotte Caucheteux}, \bibinfo{person}{Chaya Nayak}, \bibinfo{person}{Chloe Bi}, \bibinfo{person}{Chris Marra},
  \bibinfo{person}{Chris McConnell}, \bibinfo{person}{Christian Keller}, \bibinfo{person}{Christophe Touret}, \bibinfo{person}{Chunyang Wu}, \bibinfo{person}{Corinne Wong}, \bibinfo{person}{Cristian~Canton Ferrer}, \bibinfo{person}{Cyrus Nikolaidis}, \bibinfo{person}{Damien Allonsius}, \bibinfo{person}{Daniel Song}, \bibinfo{person}{Danielle Pintz}, \bibinfo{person}{Danny Livshits}, \bibinfo{person}{David Esiobu}, \bibinfo{person}{Dhruv Choudhary}, \bibinfo{person}{Dhruv Mahajan}, \bibinfo{person}{Diego Garcia{-}Olano}, \bibinfo{person}{Diego Perino}, \bibinfo{person}{Dieuwke Hupkes}, \bibinfo{person}{Egor Lakomkin}, \bibinfo{person}{Ehab AlBadawy}, \bibinfo{person}{Elina Lobanova}, \bibinfo{person}{Emily Dinan}, \bibinfo{person}{Eric~Michael Smith}, \bibinfo{person}{Filip Radenovic}, \bibinfo{person}{Frank Zhang}, \bibinfo{person}{Gabriel Synnaeve}, \bibinfo{person}{Gabrielle Lee}, \bibinfo{person}{Georgia~Lewis Anderson}, \bibinfo{person}{Graeme Nail}, \bibinfo{person}{Gr{\'{e}}goire Mialon},
  \bibinfo{person}{Guan Pang}, \bibinfo{person}{Guillem Cucurell}, \bibinfo{person}{Hailey Nguyen}, \bibinfo{person}{Hannah Korevaar}, \bibinfo{person}{Hu Xu}, \bibinfo{person}{Hugo Touvron}, \bibinfo{person}{Iliyan Zarov}, \bibinfo{person}{Imanol~Arrieta Ibarra}, \bibinfo{person}{Isabel~M. Kloumann}, \bibinfo{person}{Ishan Misra}, \bibinfo{person}{Ivan Evtimov}, \bibinfo{person}{Jade Copet}, \bibinfo{person}{Jaewon Lee}, \bibinfo{person}{Jan Geffert}, \bibinfo{person}{Jana Vranes}, \bibinfo{person}{Jason Park}, \bibinfo{person}{Jay Mahadeokar}, \bibinfo{person}{Jeet Shah}, \bibinfo{person}{Jelmer van~der Linde}, \bibinfo{person}{Jennifer Billock}, \bibinfo{person}{Jenny Hong}, \bibinfo{person}{Jenya Lee}, \bibinfo{person}{Jeremy Fu}, \bibinfo{person}{Jianfeng Chi}, \bibinfo{person}{Jianyu Huang}, \bibinfo{person}{Jiawen Liu}, \bibinfo{person}{Jie Wang}, \bibinfo{person}{Jiecao Yu}, \bibinfo{person}{Joanna Bitton}, \bibinfo{person}{Joe Spisak}, \bibinfo{person}{Jongsoo Park}, \bibinfo{person}{Joseph Rocca},
  \bibinfo{person}{Joshua Johnstun}, \bibinfo{person}{Joshua Saxe}, \bibinfo{person}{Junteng Jia}, \bibinfo{person}{Kalyan~Vasuden Alwala}, \bibinfo{person}{Kartikeya Upasani}, \bibinfo{person}{Kate Plawiak}, \bibinfo{person}{Ke Li}, \bibinfo{person}{Kenneth Heafield}, \bibinfo{person}{Kevin Stone}, {and} \bibinfo{person}{et al.}} \bibinfo{year}{2024}\natexlab{}.
\newblock \showarticletitle{The Llama 3 Herd of Models}.
\newblock \bibinfo{journal}{\emph{CoRR}}  \bibinfo{volume}{abs/2407.21783} (\bibinfo{year}{2024}).
\newblock
\urldef\tempurl%
\url{https://doi.org/10.48550/ARXIV.2407.21783}
\showDOI{\tempurl}
\showeprint[arXiv]{2407.21783}


\bibitem[Gerganov(2024)]%
        {llama.cpp}
\bibfield{author}{\bibinfo{person}{Georgi Gerganov}.} \bibinfo{year}{2024}\natexlab{}.
\newblock \bibinfo{title}{ggerganov/llama.cpp: Port of Facebook’s LLaMA model in C/C++.}
\newblock
\newblock
\newblock
\shownote{https://github.com/ggerganov/llama.cpp}.


\bibitem[He et~al\mbox{.}(2024)]%
        {he-etal-2024-chess}
\bibfield{author}{\bibinfo{person}{Junhui He}, \bibinfo{person}{Shangyu Wu}, \bibinfo{person}{Weidong Wen}, \bibinfo{person}{Chun~Jason Xue}, {and} \bibinfo{person}{Qingan Li}.} \bibinfo{year}{2024}\natexlab{}.
\newblock \showarticletitle{{CHESS}: Optimizing {LLM} Inference via Channel-Wise Thresholding and Selective Sparsification}. In \bibinfo{booktitle}{\emph{Proceedings of the 2024 Conference on Empirical Methods in Natural Language Processing}}, \bibfield{editor}{\bibinfo{person}{Yaser Al-Onaizan}, \bibinfo{person}{Mohit Bansal}, {and} \bibinfo{person}{Yun-Nung Chen}} (Eds.). \bibinfo{publisher}{Association for Computational Linguistics}, \bibinfo{address}{Miami, Florida, USA}, \bibinfo{pages}{18658--18668}.
\newblock
\urldef\tempurl%
\url{https://doi.org/10.18653/v1/2024.emnlp-main.1038}
\showDOI{\tempurl}


\bibitem[Hsieh et~al\mbox{.}(2023)]%
        {hsieh2023distilling}
\bibfield{author}{\bibinfo{person}{Cheng-Yu Hsieh}, \bibinfo{person}{Chun-Liang Li}, \bibinfo{person}{Chih-Kuan Yeh}, \bibinfo{person}{Hootan Nakhost}, \bibinfo{person}{Yasuhisa Fujii}, \bibinfo{person}{Alexander Ratner}, \bibinfo{person}{Ranjay Krishna}, \bibinfo{person}{Chen-Yu Lee}, {and} \bibinfo{person}{Tomas Pfister}.} \bibinfo{year}{2023}\natexlab{}.
\newblock \bibinfo{title}{Distilling Step-by-Step! Outperforming Larger Language Models with Less Training Data and Smaller Model Sizes}.
\newblock
\newblock
\showeprint[arxiv]{2305.02301}~[cs.CL]
\urldef\tempurl%
\url{https://arxiv.org/abs/2305.02301}
\showURL{%
\tempurl}


\bibitem[Huang et~al\mbox{.}(2024a)]%
        {24arxiv-raee}
\bibfield{author}{\bibinfo{person}{Lianming Huang}, \bibinfo{person}{Shangyu Wu}, \bibinfo{person}{Yufei Cui}, \bibinfo{person}{Ying Xiong}, \bibinfo{person}{Xue Liu}, \bibinfo{person}{Tei{-}Wei Kuo}, \bibinfo{person}{Nan Guan}, {and} \bibinfo{person}{Chun~Jason Xue}.} \bibinfo{year}{2024}\natexlab{a}.
\newblock \showarticletitle{{RAEE:} {A} Training-Free Retrieval-Augmented Early Exiting Framework for Efficient Inference}.
\newblock \bibinfo{journal}{\emph{CoRR}}  \bibinfo{volume}{abs/2405.15198} (\bibinfo{year}{2024}).
\newblock
\urldef\tempurl%
\url{https://doi.org/10.48550/ARXIV.2405.15198}
\showDOI{\tempurl}
\showeprint[arXiv]{2405.15198}


\bibitem[Huang et~al\mbox{.}(2024b)]%
        {Huang_2024}
\bibfield{author}{\bibinfo{person}{Wei Huang}, \bibinfo{person}{Xingyu Zheng}, \bibinfo{person}{Xudong Ma}, \bibinfo{person}{Haotong Qin}, \bibinfo{person}{Chengtao Lv}, \bibinfo{person}{Hong Chen}, \bibinfo{person}{Jie Luo}, \bibinfo{person}{Xiaojuan Qi}, \bibinfo{person}{Xianglong Liu}, {and} \bibinfo{person}{Michele Magno}.} \bibinfo{year}{2024}\natexlab{b}.
\newblock \showarticletitle{An empirical study of LLaMA3 quantization: from LLMs to MLLMs}.
\newblock \bibinfo{journal}{\emph{Visual Intelligence}} \bibinfo{volume}{2}, \bibinfo{number}{1} (\bibinfo{date}{Dec.} \bibinfo{year}{2024}).
\newblock
\showISSN{2731-9008}
\urldef\tempurl%
\url{https://doi.org/10.1007/s44267-024-00070-x}
\showDOI{\tempurl}


\bibitem[HuggingFace.(2022)]%
        {accelerate}
\bibfield{author}{\bibinfo{person}{HuggingFace.}} \bibinfo{year}{2022}\natexlab{}.
\newblock \bibinfo{title}{Hugging face accelerate.}
\newblock
\newblock
\newblock
\shownote{https://huggingface.co/docs/accelerate/index}.


\bibitem[Jain et~al\mbox{.}(2020)]%
        {MLSYS2020_0b816ae8}
\bibfield{author}{\bibinfo{person}{Paras Jain}, \bibinfo{person}{Ajay Jain}, \bibinfo{person}{Aniruddha Nrusimha}, \bibinfo{person}{Amir Gholami}, \bibinfo{person}{Pieter Abbeel}, \bibinfo{person}{Joseph Gonzalez}, \bibinfo{person}{Kurt Keutzer}, {and} \bibinfo{person}{Ion Stoica}.} \bibinfo{year}{2020}\natexlab{}.
\newblock \showarticletitle{Checkmate: Breaking the Memory Wall with Optimal Tensor Rematerialization}. In \bibinfo{booktitle}{\emph{Proceedings of Machine Learning and Systems}}, \bibfield{editor}{\bibinfo{person}{I.~Dhillon}, \bibinfo{person}{D.~Papailiopoulos}, {and} \bibinfo{person}{V.~Sze}} (Eds.), Vol.~\bibinfo{volume}{2}. \bibinfo{pages}{497--511}.
\newblock
\urldef\tempurl%
\url{https://proceedings.mlsys.org/paper_files/paper/2020/file/0b816ae8f06f8dd3543dc3d9ef196cab-Paper.pdf}
\showURL{%
\tempurl}


\bibitem[Ji et~al\mbox{.}(2017)]%
        {Mobile_IO}
\bibfield{author}{\bibinfo{person}{Cheng Ji}, \bibinfo{person}{Li-Pin Chang}, \bibinfo{person}{Liang Shi}, \bibinfo{person}{Congming Gao}, \bibinfo{person}{Chao Wu}, \bibinfo{person}{Yuangang Wang}, {and} \bibinfo{person}{Chun~Jason Xue}.} \bibinfo{year}{2017}\natexlab{}.
\newblock \showarticletitle{Lightweight Data Compression for Mobile Flash Storage}.
\newblock \bibinfo{journal}{\emph{ACM Trans. Embed. Comput. Syst.}} \bibinfo{volume}{16}, \bibinfo{number}{5s}, Article \bibinfo{articleno}{183} (\bibinfo{date}{Sept.} \bibinfo{year}{2017}), \bibinfo{numpages}{18}~pages.
\newblock
\showISSN{1539-9087}
\urldef\tempurl%
\url{https://doi.org/10.1145/3126511}
\showDOI{\tempurl}


\bibitem[Kaplan et~al\mbox{.}(2020)]%
        {kaplan2020scalinglaws}
\bibfield{author}{\bibinfo{person}{Jared Kaplan}, \bibinfo{person}{Sam McCandlish}, \bibinfo{person}{Tom Henighan}, \bibinfo{person}{Tom~B. Brown}, \bibinfo{person}{Benjamin Chess}, \bibinfo{person}{Rewon Child}, \bibinfo{person}{Scott Gray}, \bibinfo{person}{Alec Radford}, \bibinfo{person}{Jeffrey Wu}, {and} \bibinfo{person}{Dario Amodei}.} \bibinfo{year}{2020}\natexlab{}.
\newblock \bibinfo{title}{Scaling Laws for Neural Language Models}.
\newblock
\newblock
\showeprint[arxiv]{2001.08361}~[cs.LG]
\urldef\tempurl%
\url{https://arxiv.org/abs/2001.08361}
\showURL{%
\tempurl}


\bibitem[Kwon et~al\mbox{.}(2023)]%
        {pagedattention}
\bibfield{author}{\bibinfo{person}{Woosuk Kwon}, \bibinfo{person}{Zhuohan Li}, \bibinfo{person}{Siyuan Zhuang}, \bibinfo{person}{Ying Sheng}, \bibinfo{person}{Lianmin Zheng}, \bibinfo{person}{Cody~Hao Yu}, \bibinfo{person}{Joseph Gonzalez}, \bibinfo{person}{Hao Zhang}, {and} \bibinfo{person}{Ion Stoica}.} \bibinfo{year}{2023}\natexlab{}.
\newblock \showarticletitle{Efficient Memory Management for Large Language Model Serving with PagedAttention}. In \bibinfo{booktitle}{\emph{Proceedings of the 29th Symposium on Operating Systems Principles}} (Koblenz, Germany) \emph{(\bibinfo{series}{SOSP '23})}. \bibinfo{publisher}{Association for Computing Machinery}, \bibinfo{address}{New York, NY, USA}, \bibinfo{pages}{611–626}.
\newblock
\showISBNx{9798400702297}
\urldef\tempurl%
\url{https://doi.org/10.1145/3600006.3613165}
\showDOI{\tempurl}


\bibitem[Leviathan et~al\mbox{.}(2023)]%
        {speculative_decoding}
\bibfield{author}{\bibinfo{person}{Yaniv Leviathan}, \bibinfo{person}{Matan Kalman}, {and} \bibinfo{person}{Yossi Matias}.} \bibinfo{year}{2023}\natexlab{}.
\newblock \showarticletitle{Fast inference from transformers via speculative decoding}. In \bibinfo{booktitle}{\emph{Proceedings of the 40th International Conference on Machine Learning}} (Honolulu, Hawaii, USA) \emph{(\bibinfo{series}{ICML'23})}. \bibinfo{publisher}{JMLR.org}, Article \bibinfo{articleno}{795}, \bibinfo{numpages}{13}~pages.
\newblock


\bibitem[Lin et~al\mbox{.}(2024)]%
        {MLSYS2024_42a452cb}
\bibfield{author}{\bibinfo{person}{Ji Lin}, \bibinfo{person}{Jiaming Tang}, \bibinfo{person}{Haotian Tang}, \bibinfo{person}{Shang Yang}, \bibinfo{person}{Wei-Ming Chen}, \bibinfo{person}{Wei-Chen Wang}, \bibinfo{person}{Guangxuan Xiao}, \bibinfo{person}{Xingyu Dang}, \bibinfo{person}{Chuang Gan}, {and} \bibinfo{person}{Song Han}.} \bibinfo{year}{2024}\natexlab{}.
\newblock \showarticletitle{AWQ: Activation-aware Weight Quantization for On-Device LLM Compression and Acceleration}. In \bibinfo{booktitle}{\emph{Proceedings of Machine Learning and Systems}}, \bibfield{editor}{\bibinfo{person}{P.~Gibbons}, \bibinfo{person}{G.~Pekhimenko}, {and} \bibinfo{person}{C.~De Sa}} (Eds.), Vol.~\bibinfo{volume}{6}. \bibinfo{pages}{87--100}.
\newblock
\urldef\tempurl%
\url{https://proceedings.mlsys.org/paper_files/paper/2024/file/42a452cbafa9dd64e9ba4aa95cc1ef21-Paper-Conference.pdf}
\showURL{%
\tempurl}


\bibitem[Linux.(2024a)]%
        {cgroup}
\bibfield{author}{\bibinfo{person}{Linux.}} \bibinfo{year}{2024}\natexlab{a}.
\newblock \bibinfo{title}{cgroups - Linux control groups}.
\newblock
\newblock
\newblock
\shownote{https://man7.org/linux/man-pages/man7/cgroups.7.html}.


\bibitem[Linux.(2024b)]%
        {mmap}
\bibfield{author}{\bibinfo{person}{Linux.}} \bibinfo{year}{2024}\natexlab{b}.
\newblock \bibinfo{title}{mmap, munmap - map or unmap files or devices into memory}.
\newblock
\newblock
\newblock
\shownote{https://man7.org/linux/man-pages/man2/mmap.2.html}.


\bibitem[Linux.(2024c)]%
        {taskset}
\bibfield{author}{\bibinfo{person}{Linux.}} \bibinfo{year}{2024}\natexlab{c}.
\newblock \bibinfo{title}{taskset - set or retrieve a process's CPU affinity}.
\newblock
\newblock
\newblock
\shownote{https://man7.org/linux/man-pages/man1/taskset.1.html}.


\bibitem[Liu et~al\mbox{.}(2023a)]%
        {pmlr-v202-liu23am}
\bibfield{author}{\bibinfo{person}{Zichang Liu}, \bibinfo{person}{Jue Wang}, \bibinfo{person}{Tri Dao}, \bibinfo{person}{Tianyi Zhou}, \bibinfo{person}{Binhang Yuan}, \bibinfo{person}{Zhao Song}, \bibinfo{person}{Anshumali Shrivastava}, \bibinfo{person}{Ce Zhang}, \bibinfo{person}{Yuandong Tian}, \bibinfo{person}{Christopher Re}, {and} \bibinfo{person}{Beidi Chen}.} \bibinfo{year}{2023}\natexlab{a}.
\newblock \showarticletitle{Deja Vu: Contextual Sparsity for Efficient {LLM}s at Inference Time}. In \bibinfo{booktitle}{\emph{Proceedings of the 40th International Conference on Machine Learning}} \emph{(\bibinfo{series}{Proceedings of Machine Learning Research}, Vol.~\bibinfo{volume}{202})}, \bibfield{editor}{\bibinfo{person}{Andreas Krause}, \bibinfo{person}{Emma Brunskill}, \bibinfo{person}{Kyunghyun Cho}, \bibinfo{person}{Barbara Engelhardt}, \bibinfo{person}{Sivan Sabato}, {and} \bibinfo{person}{Jonathan Scarlett}} (Eds.). \bibinfo{publisher}{PMLR}, \bibinfo{pages}{22137--22176}.
\newblock
\urldef\tempurl%
\url{https://proceedings.mlr.press/v202/liu23am.html}
\showURL{%
\tempurl}


\bibitem[Liu et~al\mbox{.}(2023b)]%
        {DejaVu}
\bibfield{author}{\bibinfo{person}{Zichang Liu}, \bibinfo{person}{Jue Wang}, \bibinfo{person}{Tri Dao}, \bibinfo{person}{Tianyi Zhou}, \bibinfo{person}{Binhang Yuan}, \bibinfo{person}{Zhao Song}, \bibinfo{person}{Anshumali Shrivastava}, \bibinfo{person}{Ce Zhang}, \bibinfo{person}{Yuandong Tian}, \bibinfo{person}{Christopher Re}, {and} \bibinfo{person}{Beidi Chen}.} \bibinfo{year}{2023}\natexlab{b}.
\newblock \showarticletitle{Deja Vu: Contextual Sparsity for Efficient {LLM}s at Inference Time}. In \bibinfo{booktitle}{\emph{Proceedings of the 40th International Conference on Machine Learning}} \emph{(\bibinfo{series}{Proceedings of Machine Learning Research}, Vol.~\bibinfo{volume}{202})}, \bibfield{editor}{\bibinfo{person}{Andreas Krause}, \bibinfo{person}{Emma Brunskill}, \bibinfo{person}{Kyunghyun Cho}, \bibinfo{person}{Barbara Engelhardt}, \bibinfo{person}{Sivan Sabato}, {and} \bibinfo{person}{Jonathan Scarlett}} (Eds.). \bibinfo{publisher}{PMLR}, \bibinfo{pages}{22137--22176}.
\newblock
\urldef\tempurl%
\url{https://proceedings.mlr.press/v202/liu23am.html}
\showURL{%
\tempurl}


\bibitem[Lyu et~al\mbox{.}(2024)]%
        {lyu2024llmrec}
\bibfield{author}{\bibinfo{person}{Hanjia Lyu}, \bibinfo{person}{Song Jiang}, \bibinfo{person}{Hanqing Zeng}, \bibinfo{person}{Yinglong Xia}, \bibinfo{person}{Qifan Wang}, \bibinfo{person}{Si Zhang}, \bibinfo{person}{Ren Chen}, \bibinfo{person}{Christopher Leung}, \bibinfo{person}{Jiajie Tang}, {and} \bibinfo{person}{Jiebo Luo}.} \bibinfo{year}{2024}\natexlab{}.
\newblock \bibinfo{title}{LLM-Rec: Personalized Recommendation via Prompting Large Language Models}.
\newblock
\newblock
\showeprint[arxiv]{2307.15780}~[cs.CL]
\urldef\tempurl%
\url{https://arxiv.org/abs/2307.15780}
\showURL{%
\tempurl}


\bibitem[Mao et~al\mbox{.}(2024)]%
        {mao2024compressibility}
\bibfield{author}{\bibinfo{person}{Yu Mao}, \bibinfo{person}{Weilan Wang}, \bibinfo{person}{Hongchao Du}, \bibinfo{person}{Nan Guan}, {and} \bibinfo{person}{Chun~Jason Xue}.} \bibinfo{year}{2024}\natexlab{}.
\newblock \bibinfo{title}{On the Compressibility of Quantized Large Language Models}.
\newblock
\newblock
\showeprint[arxiv]{2403.01384}~[cs.LG]
\urldef\tempurl%
\url{https://arxiv.org/abs/2403.01384}
\showURL{%
\tempurl}


\bibitem[NVIDIA.(2023)]%
        {FasterTransformer}
\bibfield{author}{\bibinfo{person}{NVIDIA.}} \bibinfo{year}{2023}\natexlab{}.
\newblock \bibinfo{title}{FasterTransformer}.
\newblock
\newblock
\newblock
\shownote{https://developer.nvidia.com/nvidia-triton-inference-server}.


\bibitem[OpenAI et~al\mbox{.}(2024)]%
        {openai2024openaio1card}
\bibfield{author}{\bibinfo{person}{OpenAI}, \bibinfo{person}{:}, \bibinfo{person}{Aaron Jaech}, \bibinfo{person}{Adam Kalai}, \bibinfo{person}{Adam Lerer}, \bibinfo{person}{Adam Richardson}, \bibinfo{person}{Ahmed El-Kishky}, \bibinfo{person}{Aiden Low}, \bibinfo{person}{Alec Helyar}, \bibinfo{person}{Aleksander Madry}, \bibinfo{person}{Alex Beutel}, \bibinfo{person}{Alex Carney}, \bibinfo{person}{Alex Iftimie}, \bibinfo{person}{Alex Karpenko}, \bibinfo{person}{Alex~Tachard Passos}, \bibinfo{person}{Alexander Neitz}, \bibinfo{person}{Alexander Prokofiev}, \bibinfo{person}{Alexander Wei}, \bibinfo{person}{Allison Tam}, \bibinfo{person}{Ally Bennett}, \bibinfo{person}{Ananya Kumar}, \bibinfo{person}{Andre Saraiva}, \bibinfo{person}{Andrea Vallone}, \bibinfo{person}{Andrew Duberstein}, \bibinfo{person}{Andrew Kondrich}, \bibinfo{person}{Andrey Mishchenko}, \bibinfo{person}{Andy Applebaum}, \bibinfo{person}{Angela Jiang}, \bibinfo{person}{Ashvin Nair}, \bibinfo{person}{Barret Zoph}, \bibinfo{person}{Behrooz
  Ghorbani}, \bibinfo{person}{Ben Rossen}, \bibinfo{person}{Benjamin Sokolowsky}, \bibinfo{person}{Boaz Barak}, \bibinfo{person}{Bob McGrew}, \bibinfo{person}{Borys Minaiev}, \bibinfo{person}{Botao Hao}, \bibinfo{person}{Bowen Baker}, \bibinfo{person}{Brandon Houghton}, \bibinfo{person}{Brandon McKinzie}, \bibinfo{person}{Brydon Eastman}, \bibinfo{person}{Camillo Lugaresi}, \bibinfo{person}{Cary Bassin}, \bibinfo{person}{Cary Hudson}, \bibinfo{person}{Chak~Ming Li}, \bibinfo{person}{Charles de Bourcy}, \bibinfo{person}{Chelsea Voss}, \bibinfo{person}{Chen Shen}, \bibinfo{person}{Chong Zhang}, \bibinfo{person}{Chris Koch}, \bibinfo{person}{Chris Orsinger}, \bibinfo{person}{Christopher Hesse}, \bibinfo{person}{Claudia Fischer}, \bibinfo{person}{Clive Chan}, \bibinfo{person}{Dan Roberts}, \bibinfo{person}{Daniel Kappler}, \bibinfo{person}{Daniel Levy}, \bibinfo{person}{Daniel Selsam}, \bibinfo{person}{David Dohan}, \bibinfo{person}{David Farhi}, \bibinfo{person}{David Mely}, \bibinfo{person}{David Robinson},
  \bibinfo{person}{Dimitris Tsipras}, \bibinfo{person}{Doug Li}, \bibinfo{person}{Dragos Oprica}, \bibinfo{person}{Eben Freeman}, \bibinfo{person}{Eddie Zhang}, \bibinfo{person}{Edmund Wong}, \bibinfo{person}{Elizabeth Proehl}, \bibinfo{person}{Enoch Cheung}, \bibinfo{person}{Eric Mitchell}, \bibinfo{person}{Eric Wallace}, \bibinfo{person}{Erik Ritter}, \bibinfo{person}{Evan Mays}, \bibinfo{person}{Fan Wang}, \bibinfo{person}{Felipe~Petroski Such}, \bibinfo{person}{Filippo Raso}, \bibinfo{person}{Florencia Leoni}, \bibinfo{person}{Foivos Tsimpourlas}, \bibinfo{person}{Francis Song}, \bibinfo{person}{Fred von Lohmann}, \bibinfo{person}{Freddie Sulit}, \bibinfo{person}{Geoff Salmon}, \bibinfo{person}{Giambattista Parascandolo}, \bibinfo{person}{Gildas Chabot}, \bibinfo{person}{Grace Zhao}, \bibinfo{person}{Greg Brockman}, \bibinfo{person}{Guillaume Leclerc}, \bibinfo{person}{Hadi Salman}, \bibinfo{person}{Haiming Bao}, \bibinfo{person}{Hao Sheng}, \bibinfo{person}{Hart Andrin}, \bibinfo{person}{Hessam
  Bagherinezhad}, \bibinfo{person}{Hongyu Ren}, \bibinfo{person}{Hunter Lightman}, \bibinfo{person}{Hyung~Won Chung}, \bibinfo{person}{Ian Kivlichan}, \bibinfo{person}{Ian O'Connell}, \bibinfo{person}{Ian Osband}, \bibinfo{person}{Ignasi~Clavera Gilaberte}, \bibinfo{person}{Ilge Akkaya}, \bibinfo{person}{Ilya Kostrikov}, \bibinfo{person}{Ilya Sutskever}, \bibinfo{person}{Irina Kofman}, \bibinfo{person}{Jakub Pachocki}, \bibinfo{person}{James Lennon}, \bibinfo{person}{Jason Wei}, \bibinfo{person}{Jean Harb}, \bibinfo{person}{Jerry Twore}, \bibinfo{person}{Jiacheng Feng}, \bibinfo{person}{Jiahui Yu}, \bibinfo{person}{Jiayi Weng}, \bibinfo{person}{Jie Tang}, \bibinfo{person}{Jieqi Yu}, \bibinfo{person}{Joaquin~Quiñonero Candela}, \bibinfo{person}{Joe Palermo}, \bibinfo{person}{Joel Parish}, \bibinfo{person}{Johannes Heidecke}, \bibinfo{person}{John Hallman}, \bibinfo{person}{John Rizzo}, \bibinfo{person}{Jonathan Gordon}, \bibinfo{person}{Jonathan Uesato}, \bibinfo{person}{Jonathan Ward}, \bibinfo{person}{Joost
  Huizinga}, \bibinfo{person}{Julie Wang}, \bibinfo{person}{Kai Chen}, \bibinfo{person}{Kai Xiao}, \bibinfo{person}{Karan Singhal}, \bibinfo{person}{Karina Nguyen}, \bibinfo{person}{Karl Cobbe}, \bibinfo{person}{Katy Shi}, \bibinfo{person}{Kayla Wood}, \bibinfo{person}{Kendra Rimbach}, \bibinfo{person}{Keren Gu-Lemberg}, \bibinfo{person}{Kevin Liu}, \bibinfo{person}{Kevin Lu}, \bibinfo{person}{Kevin Stone}, \bibinfo{person}{Kevin Yu}, \bibinfo{person}{Lama Ahmad}, \bibinfo{person}{Lauren Yang}, \bibinfo{person}{Leo Liu}, \bibinfo{person}{Leon Maksin}, \bibinfo{person}{Leyton Ho}, \bibinfo{person}{Liam Fedus}, \bibinfo{person}{Lilian Weng}, \bibinfo{person}{Linden Li}, \bibinfo{person}{Lindsay McCallum}, \bibinfo{person}{Lindsey Held}, \bibinfo{person}{Lorenz Kuhn}, \bibinfo{person}{Lukas Kondraciuk}, \bibinfo{person}{Lukasz Kaiser}, \bibinfo{person}{Luke Metz}, \bibinfo{person}{Madelaine Boyd}, \bibinfo{person}{Maja Trebacz}, \bibinfo{person}{Manas Joglekar}, \bibinfo{person}{Mark Chen},
  \bibinfo{person}{Marko Tintor}, \bibinfo{person}{Mason Meyer}, \bibinfo{person}{Matt Jones}, \bibinfo{person}{Matt Kaufer}, \bibinfo{person}{Max Schwarzer}, \bibinfo{person}{Meghan Shah}, \bibinfo{person}{Mehmet Yatbaz}, \bibinfo{person}{Melody~Y. Guan}, \bibinfo{person}{Mengyuan Xu}, \bibinfo{person}{Mengyuan Yan}, \bibinfo{person}{Mia Glaese}, \bibinfo{person}{Mianna Chen}, \bibinfo{person}{Michael Lampe}, \bibinfo{person}{Michael Malek}, \bibinfo{person}{Michele Wang}, \bibinfo{person}{Michelle Fradin}, \bibinfo{person}{Mike McClay}, \bibinfo{person}{Mikhail Pavlov}, \bibinfo{person}{Miles Wang}, \bibinfo{person}{Mingxuan Wang}, \bibinfo{person}{Mira Murati}, \bibinfo{person}{Mo Bavarian}, \bibinfo{person}{Mostafa Rohaninejad}, \bibinfo{person}{Nat McAleese}, \bibinfo{person}{Neil Chowdhury}, \bibinfo{person}{Neil Chowdhury}, \bibinfo{person}{Nick Ryder}, \bibinfo{person}{Nikolas Tezak}, \bibinfo{person}{Noam Brown}, \bibinfo{person}{Ofir Nachum}, \bibinfo{person}{Oleg Boiko}, \bibinfo{person}{Oleg
  Murk}, \bibinfo{person}{Olivia Watkins}, \bibinfo{person}{Patrick Chao}, \bibinfo{person}{Paul Ashbourne}, \bibinfo{person}{Pavel Izmailov}, \bibinfo{person}{Peter Zhokhov}, \bibinfo{person}{Rachel Dias}, \bibinfo{person}{Rahul Arora}, \bibinfo{person}{Randall Lin}, \bibinfo{person}{Rapha~Gontijo Lopes}, \bibinfo{person}{Raz Gaon}, \bibinfo{person}{Reah Miyara}, \bibinfo{person}{Reimar Leike}, \bibinfo{person}{Renny Hwang}, \bibinfo{person}{Rhythm Garg}, \bibinfo{person}{Robin Brown}, \bibinfo{person}{Roshan James}, \bibinfo{person}{Rui Shu}, \bibinfo{person}{Ryan Cheu}, \bibinfo{person}{Ryan Greene}, \bibinfo{person}{Saachi Jain}, \bibinfo{person}{Sam Altman}, \bibinfo{person}{Sam Toizer}, \bibinfo{person}{Sam Toyer}, \bibinfo{person}{Samuel Miserendino}, \bibinfo{person}{Sandhini Agarwal}, \bibinfo{person}{Santiago Hernandez}, \bibinfo{person}{Sasha Baker}, \bibinfo{person}{Scott McKinney}, \bibinfo{person}{Scottie Yan}, \bibinfo{person}{Shengjia Zhao}, \bibinfo{person}{Shengli Hu},
  \bibinfo{person}{Shibani Santurkar}, \bibinfo{person}{Shraman~Ray Chaudhuri}, \bibinfo{person}{Shuyuan Zhang}, \bibinfo{person}{Siyuan Fu}, \bibinfo{person}{Spencer Papay}, \bibinfo{person}{Steph Lin}, \bibinfo{person}{Suchir Balaji}, \bibinfo{person}{Suvansh Sanjeev}, \bibinfo{person}{Szymon Sidor}, \bibinfo{person}{Tal Broda}, \bibinfo{person}{Aidan Clark}, \bibinfo{person}{Tao Wang}, \bibinfo{person}{Taylor Gordon}, \bibinfo{person}{Ted Sanders}, \bibinfo{person}{Tejal Patwardhan}, \bibinfo{person}{Thibault Sottiaux}, \bibinfo{person}{Thomas Degry}, \bibinfo{person}{Thomas Dimson}, \bibinfo{person}{Tianhao Zheng}, \bibinfo{person}{Timur Garipov}, \bibinfo{person}{Tom Stasi}, \bibinfo{person}{Trapit Bansal}, \bibinfo{person}{Trevor Creech}, \bibinfo{person}{Troy Peterson}, \bibinfo{person}{Tyna Eloundou}, \bibinfo{person}{Valerie Qi}, \bibinfo{person}{Vineet Kosaraju}, \bibinfo{person}{Vinnie Monaco}, \bibinfo{person}{Vitchyr Pong}, \bibinfo{person}{Vlad Fomenko}, \bibinfo{person}{Weiyi Zheng},
  \bibinfo{person}{Wenda Zhou}, \bibinfo{person}{Wes McCabe}, \bibinfo{person}{Wojciech Zaremba}, \bibinfo{person}{Yann Dubois}, \bibinfo{person}{Yinghai Lu}, \bibinfo{person}{Yining Chen}, \bibinfo{person}{Young Cha}, \bibinfo{person}{Yu Bai}, \bibinfo{person}{Yuchen He}, \bibinfo{person}{Yuchen Zhang}, \bibinfo{person}{Yunyun Wang}, \bibinfo{person}{Zheng Shao}, {and} \bibinfo{person}{Zhuohan Li}.} \bibinfo{year}{2024}\natexlab{}.
\newblock \bibinfo{title}{OpenAI o1 System Card}.
\newblock
\newblock
\showeprint[arxiv]{2412.16720}~[cs.AI]
\urldef\tempurl%
\url{https://arxiv.org/abs/2412.16720}
\showURL{%
\tempurl}


\bibitem[Patil et~al\mbox{.}(2022)]%
        {pmlr-v162-patil22b}
\bibfield{author}{\bibinfo{person}{Shishir~G. Patil}, \bibinfo{person}{Paras Jain}, \bibinfo{person}{Prabal Dutta}, \bibinfo{person}{Ion Stoica}, {and} \bibinfo{person}{Joseph Gonzalez}.} \bibinfo{year}{2022}\natexlab{}.
\newblock \showarticletitle{{POET}: Training Neural Networks on Tiny Devices with Integrated Rematerialization and Paging}. In \bibinfo{booktitle}{\emph{Proceedings of the 39th International Conference on Machine Learning}} \emph{(\bibinfo{series}{Proceedings of Machine Learning Research}, Vol.~\bibinfo{volume}{162})}, \bibfield{editor}{\bibinfo{person}{Kamalika Chaudhuri}, \bibinfo{person}{Stefanie Jegelka}, \bibinfo{person}{Le~Song}, \bibinfo{person}{Csaba Szepesvari}, \bibinfo{person}{Gang Niu}, {and} \bibinfo{person}{Sivan Sabato}} (Eds.). \bibinfo{publisher}{PMLR}, \bibinfo{pages}{17573--17583}.
\newblock
\urldef\tempurl%
\url{https://proceedings.mlr.press/v162/patil22b.html}
\showURL{%
\tempurl}


\bibitem[Pope et~al\mbox{.}(2022)]%
        {pope2022efficientlyscalingtransformerinference}
\bibfield{author}{\bibinfo{person}{Reiner Pope}, \bibinfo{person}{Sholto Douglas}, \bibinfo{person}{Aakanksha Chowdhery}, \bibinfo{person}{Jacob Devlin}, \bibinfo{person}{James Bradbury}, \bibinfo{person}{Anselm Levskaya}, \bibinfo{person}{Jonathan Heek}, \bibinfo{person}{Kefan Xiao}, \bibinfo{person}{Shivani Agrawal}, {and} \bibinfo{person}{Jeff Dean}.} \bibinfo{year}{2022}\natexlab{}.
\newblock \bibinfo{title}{Efficiently Scaling Transformer Inference}.
\newblock
\newblock
\showeprint[arxiv]{2211.05102}~[cs.LG]
\urldef\tempurl%
\url{https://arxiv.org/abs/2211.05102}
\showURL{%
\tempurl}


\bibitem[PrivateGPT(2023)]%
        {PrivateGPT}
PrivateGPT \bibinfo{year}{2023}\natexlab{}.
\newblock \bibinfo{title}{PrivateGPT}.
\newblock
\newblock
\newblock
\shownote{https://github.com/zylon-ai/private-gpt}.


\bibitem[Ren et~al\mbox{.}(2021)]%
        {zero-offload}
\bibfield{author}{\bibinfo{person}{Jie Ren}, \bibinfo{person}{Samyam Rajbhandari}, \bibinfo{person}{Reza~Yazdani Aminabadi}, \bibinfo{person}{Olatunji Ruwase}, \bibinfo{person}{Shuangyan Yang}, \bibinfo{person}{Minjia Zhang}, \bibinfo{person}{Dong Li}, {and} \bibinfo{person}{Yuxiong He}.} \bibinfo{year}{2021}\natexlab{}.
\newblock \showarticletitle{{ZeRO-Offload}: Democratizing {Billion-Scale} Model Training}. In \bibinfo{booktitle}{\emph{2021 USENIX Annual Technical Conference (USENIX ATC 21)}}. \bibinfo{publisher}{USENIX Association}, \bibinfo{pages}{551--564}.
\newblock
\showISBNx{978-1-939133-23-6}
\urldef\tempurl%
\url{https://www.usenix.org/conference/atc21/presentation/ren-jie}
\showURL{%
\tempurl}


\bibitem[Sheng et~al\mbox{.}(2023a)]%
        {23icml-flexgen}
\bibfield{author}{\bibinfo{person}{Ying Sheng}, \bibinfo{person}{Lianmin Zheng}, \bibinfo{person}{Binhang Yuan}, \bibinfo{person}{Zhuohan Li}, \bibinfo{person}{Max Ryabinin}, \bibinfo{person}{Beidi Chen}, \bibinfo{person}{Percy Liang}, \bibinfo{person}{Christopher R{\'{e}}}, \bibinfo{person}{Ion Stoica}, {and} \bibinfo{person}{Ce Zhang}.} \bibinfo{year}{2023}\natexlab{a}.
\newblock \showarticletitle{FlexGen: High-Throughput Generative Inference of Large Language Models with a Single {GPU}}. In \bibinfo{booktitle}{\emph{International Conference on Machine Learning, {ICML} 2023, 23-29 July 2023, Honolulu, Hawaii, {USA}}} \emph{(\bibinfo{series}{Proceedings of Machine Learning Research}, Vol.~\bibinfo{volume}{202})}, \bibfield{editor}{\bibinfo{person}{Andreas Krause}, \bibinfo{person}{Emma Brunskill}, \bibinfo{person}{Kyunghyun Cho}, \bibinfo{person}{Barbara Engelhardt}, \bibinfo{person}{Sivan Sabato}, {and} \bibinfo{person}{Jonathan Scarlett}} (Eds.). \bibinfo{publisher}{{PMLR}}, \bibinfo{pages}{31094--31116}.
\newblock
\urldef\tempurl%
\url{https://proceedings.mlr.press/v202/sheng23a.html}
\showURL{%
\tempurl}


\bibitem[Sheng et~al\mbox{.}(2023b)]%
        {flexgen}
\bibfield{author}{\bibinfo{person}{Ying Sheng}, \bibinfo{person}{Lianmin Zheng}, \bibinfo{person}{Binhang Yuan}, \bibinfo{person}{Zhuohan Li}, \bibinfo{person}{Max Ryabinin}, \bibinfo{person}{Beidi Chen}, \bibinfo{person}{Percy Liang}, \bibinfo{person}{Christopher Re}, \bibinfo{person}{Ion Stoica}, {and} \bibinfo{person}{Ce Zhang}.} \bibinfo{year}{2023}\natexlab{b}.
\newblock \showarticletitle{{F}lex{G}en: High-Throughput Generative Inference of Large Language Models with a Single {GPU}}. In \bibinfo{booktitle}{\emph{Proceedings of the 40th International Conference on Machine Learning}} \emph{(\bibinfo{series}{Proceedings of Machine Learning Research}, Vol.~\bibinfo{volume}{202})}, \bibfield{editor}{\bibinfo{person}{Andreas Krause}, \bibinfo{person}{Emma Brunskill}, \bibinfo{person}{Kyunghyun Cho}, \bibinfo{person}{Barbara Engelhardt}, \bibinfo{person}{Sivan Sabato}, {and} \bibinfo{person}{Jonathan Scarlett}} (Eds.). \bibinfo{publisher}{PMLR}, \bibinfo{pages}{31094--31116}.
\newblock
\urldef\tempurl%
\url{https://proceedings.mlr.press/v202/sheng23a.html}
\showURL{%
\tempurl}


\bibitem[Song et~al\mbox{.}(2024)]%
        {powerinfer}
\bibfield{author}{\bibinfo{person}{Yixin Song}, \bibinfo{person}{Zeyu Mi}, \bibinfo{person}{Haotong Xie}, {and} \bibinfo{person}{Haibo Chen}.} \bibinfo{year}{2024}\natexlab{}.
\newblock \showarticletitle{PowerInfer: Fast Large Language Model Serving with a Consumer-grade GPU}. In \bibinfo{booktitle}{\emph{Proceedings of the ACM SIGOPS 30th Symposium on Operating Systems Principles}} (Austin, TX, USA) \emph{(\bibinfo{series}{SOSP '24})}. \bibinfo{publisher}{Association for Computing Machinery}, \bibinfo{address}{New York, NY, USA}, \bibinfo{pages}{590–606}.
\newblock
\showISBNx{9798400712517}
\urldef\tempurl%
\url{https://doi.org/10.1145/3694715.3695964}
\showDOI{\tempurl}


\bibitem[Touvron et~al\mbox{.}(2023a)]%
        {23arxiv-llama}
\bibfield{author}{\bibinfo{person}{Hugo Touvron}, \bibinfo{person}{Thibaut Lavril}, \bibinfo{person}{Gautier Izacard}, \bibinfo{person}{Xavier Martinet}, \bibinfo{person}{Marie{-}Anne Lachaux}, \bibinfo{person}{Timoth{\'{e}}e Lacroix}, \bibinfo{person}{Baptiste Rozi{\`{e}}re}, \bibinfo{person}{Naman Goyal}, \bibinfo{person}{Eric Hambro}, \bibinfo{person}{Faisal Azhar}, \bibinfo{person}{Aur{\'{e}}lien Rodriguez}, \bibinfo{person}{Armand Joulin}, \bibinfo{person}{Edouard Grave}, {and} \bibinfo{person}{Guillaume Lample}.} \bibinfo{year}{2023}\natexlab{a}.
\newblock \showarticletitle{LLaMA: Open and Efficient Foundation Language Models}.
\newblock \bibinfo{journal}{\emph{CoRR}}  \bibinfo{volume}{abs/2302.13971} (\bibinfo{year}{2023}).
\newblock
\urldef\tempurl%
\url{https://doi.org/10.48550/ARXIV.2302.13971}
\showDOI{\tempurl}
\showeprint[arXiv]{2302.13971}


\bibitem[Touvron et~al\mbox{.}(2023b)]%
        {23arxiv-llama-2}
\bibfield{author}{\bibinfo{person}{Hugo Touvron}, \bibinfo{person}{Louis Martin}, \bibinfo{person}{Kevin Stone}, \bibinfo{person}{Peter Albert}, \bibinfo{person}{Amjad Almahairi}, \bibinfo{person}{Yasmine Babaei}, \bibinfo{person}{Nikolay Bashlykov}, \bibinfo{person}{Soumya Batra}, \bibinfo{person}{Prajjwal Bhargava}, \bibinfo{person}{Shruti Bhosale}, \bibinfo{person}{Dan Bikel}, \bibinfo{person}{Lukas Blecher}, \bibinfo{person}{Cristian Canton{-}Ferrer}, \bibinfo{person}{Moya Chen}, \bibinfo{person}{Guillem Cucurull}, \bibinfo{person}{David Esiobu}, \bibinfo{person}{Jude Fernandes}, \bibinfo{person}{Jeremy Fu}, \bibinfo{person}{Wenyin Fu}, \bibinfo{person}{Brian Fuller}, \bibinfo{person}{Cynthia Gao}, \bibinfo{person}{Vedanuj Goswami}, \bibinfo{person}{Naman Goyal}, \bibinfo{person}{Anthony Hartshorn}, \bibinfo{person}{Saghar Hosseini}, \bibinfo{person}{Rui Hou}, \bibinfo{person}{Hakan Inan}, \bibinfo{person}{Marcin Kardas}, \bibinfo{person}{Viktor Kerkez}, \bibinfo{person}{Madian Khabsa},
  \bibinfo{person}{Isabel Kloumann}, \bibinfo{person}{Artem Korenev}, \bibinfo{person}{Punit~Singh Koura}, \bibinfo{person}{Marie{-}Anne Lachaux}, \bibinfo{person}{Thibaut Lavril}, \bibinfo{person}{Jenya Lee}, \bibinfo{person}{Diana Liskovich}, \bibinfo{person}{Yinghai Lu}, \bibinfo{person}{Yuning Mao}, \bibinfo{person}{Xavier Martinet}, \bibinfo{person}{Todor Mihaylov}, \bibinfo{person}{Pushkar Mishra}, \bibinfo{person}{Igor Molybog}, \bibinfo{person}{Yixin Nie}, \bibinfo{person}{Andrew Poulton}, \bibinfo{person}{Jeremy Reizenstein}, \bibinfo{person}{Rashi Rungta}, \bibinfo{person}{Kalyan Saladi}, \bibinfo{person}{Alan Schelten}, \bibinfo{person}{Ruan Silva}, \bibinfo{person}{Eric~Michael Smith}, \bibinfo{person}{Ranjan Subramanian}, \bibinfo{person}{Xiaoqing~Ellen Tan}, \bibinfo{person}{Binh Tang}, \bibinfo{person}{Ross Taylor}, \bibinfo{person}{Adina Williams}, \bibinfo{person}{Jian~Xiang Kuan}, \bibinfo{person}{Puxin Xu}, \bibinfo{person}{Zheng Yan}, \bibinfo{person}{Iliyan Zarov}, \bibinfo{person}{Yuchen
  Zhang}, \bibinfo{person}{Angela Fan}, \bibinfo{person}{Melanie Kambadur}, \bibinfo{person}{Sharan Narang}, \bibinfo{person}{Aur{\'{e}}lien Rodriguez}, \bibinfo{person}{Robert Stojnic}, \bibinfo{person}{Sergey Edunov}, {and} \bibinfo{person}{Thomas Scialom}.} \bibinfo{year}{2023}\natexlab{b}.
\newblock \showarticletitle{Llama 2: Open Foundation and Fine-Tuned Chat Models}.
\newblock \bibinfo{journal}{\emph{CoRR}}  \bibinfo{volume}{abs/2307.09288} (\bibinfo{year}{2023}).
\newblock
\urldef\tempurl%
\url{https://doi.org/10.48550/ARXIV.2307.09288}
\showDOI{\tempurl}
\showeprint[arXiv]{2307.09288}


\bibitem[Vaswani et~al\mbox{.}(2017)]%
        {17nips-trans}
\bibfield{author}{\bibinfo{person}{Ashish Vaswani}, \bibinfo{person}{Noam Shazeer}, \bibinfo{person}{Niki Parmar}, \bibinfo{person}{Jakob Uszkoreit}, \bibinfo{person}{Llion Jones}, \bibinfo{person}{Aidan~N. Gomez}, \bibinfo{person}{Lukasz Kaiser}, {and} \bibinfo{person}{Illia Polosukhin}.} \bibinfo{year}{2017}\natexlab{}.
\newblock \showarticletitle{Attention is All you Need}. In \bibinfo{booktitle}{\emph{Advances in Neural Information Processing Systems 30: Annual Conference on Neural Information Processing Systems 2017, December 4-9, 2017, Long Beach, CA, {USA}}}, \bibfield{editor}{\bibinfo{person}{Isabelle Guyon}, \bibinfo{person}{Ulrike von Luxburg}, \bibinfo{person}{Samy Bengio}, \bibinfo{person}{Hanna~M. Wallach}, \bibinfo{person}{Rob Fergus}, \bibinfo{person}{S.~V.~N. Vishwanathan}, {and} \bibinfo{person}{Roman Garnett}} (Eds.). \bibinfo{pages}{5998--6008}.
\newblock
\urldef\tempurl%
\url{https://proceedings.neurips.cc/paper/2017/hash/3f5ee243547dee91fbd053c1c4a845aa-Abstract.html}
\showURL{%
\tempurl}


\bibitem[Wang et~al\mbox{.}(2018)]%
        {Superneurons}
\bibfield{author}{\bibinfo{person}{Linnan Wang}, \bibinfo{person}{Jinmian Ye}, \bibinfo{person}{Yiyang Zhao}, \bibinfo{person}{Wei Wu}, \bibinfo{person}{Ang Li}, \bibinfo{person}{Shuaiwen~Leon Song}, \bibinfo{person}{Zenglin Xu}, {and} \bibinfo{person}{Tim Kraska}.} \bibinfo{year}{2018}\natexlab{}.
\newblock \showarticletitle{Superneurons: dynamic GPU memory management for training deep neural networks}.
\newblock \bibinfo{journal}{\emph{SIGPLAN Not.}} \bibinfo{volume}{53}, \bibinfo{number}{1} (\bibinfo{date}{Feb.} \bibinfo{year}{2018}), \bibinfo{pages}{41–53}.
\newblock
\showISSN{0362-1340}
\urldef\tempurl%
\url{https://doi.org/10.1145/3200691.3178491}
\showDOI{\tempurl}


\bibitem[Wang et~al\mbox{.}(2024)]%
        {wang-etal-2024-compression}
\bibfield{author}{\bibinfo{person}{Weilan Wang}, \bibinfo{person}{Yu Mao}, \bibinfo{person}{Tang Dongdong}, \bibinfo{person}{Du Hongchao}, \bibinfo{person}{Nan Guan}, {and} \bibinfo{person}{Chun~Jason Xue}.} \bibinfo{year}{2024}\natexlab{}.
\newblock \showarticletitle{When Compression Meets Model Compression: Memory-Efficient Double Compression for Large Language Models}. In \bibinfo{booktitle}{\emph{Findings of the Association for Computational Linguistics: EMNLP 2024}}, \bibfield{editor}{\bibinfo{person}{Yaser Al-Onaizan}, \bibinfo{person}{Mohit Bansal}, {and} \bibinfo{person}{Yun-Nung Chen}} (Eds.). \bibinfo{publisher}{Association for Computational Linguistics}, \bibinfo{address}{Miami, Florida, USA}, \bibinfo{pages}{16973--16983}.
\newblock
\urldef\tempurl%
\url{https://doi.org/10.18653/v1/2024.findings-emnlp.988}
\showDOI{\tempurl}


\bibitem[Wang et~al\mbox{.}(2021)]%
        {wang-etal-2021-lightseq}
\bibfield{author}{\bibinfo{person}{Xiaohui Wang}, \bibinfo{person}{Ying Xiong}, \bibinfo{person}{Yang Wei}, \bibinfo{person}{Mingxuan Wang}, {and} \bibinfo{person}{Lei Li}.} \bibinfo{year}{2021}\natexlab{}.
\newblock \showarticletitle{{L}ight{S}eq: A High Performance Inference Library for Transformers}. In \bibinfo{booktitle}{\emph{Proceedings of the 2021 Conference of the North American Chapter of the Association for Computational Linguistics: Human Language Technologies: Industry Papers}}, \bibfield{editor}{\bibinfo{person}{Young-bum Kim}, \bibinfo{person}{Yunyao Li}, {and} \bibinfo{person}{Owen Rambow}} (Eds.). \bibinfo{publisher}{Association for Computational Linguistics}, \bibinfo{address}{Online}, \bibinfo{pages}{113--120}.
\newblock
\urldef\tempurl%
\url{https://doi.org/10.18653/v1/2021.naacl-industry.15}
\showDOI{\tempurl}


\bibitem[Wu et~al\mbox{.}(2025)]%
        {wu2025evoprobustllminference}
\bibfield{author}{\bibinfo{person}{Shangyu Wu}, \bibinfo{person}{Hongchao Du}, \bibinfo{person}{Ying Xiong}, \bibinfo{person}{Shuai Chen}, \bibinfo{person}{Tei wei Kuo}, \bibinfo{person}{Nan Guan}, {and} \bibinfo{person}{Chun~Jason Xue}.} \bibinfo{year}{2025}\natexlab{}.
\newblock \bibinfo{title}{EvoP: Robust LLM Inference via Evolutionary Pruning}.
\newblock
\newblock
\showeprint[arxiv]{2502.14910}~[cs.CL]
\urldef\tempurl%
\url{https://arxiv.org/abs/2502.14910}
\showURL{%
\tempurl}


\bibitem[Xia et~al\mbox{.}(2023)]%
        {xia2023flashllm}
\bibfield{author}{\bibinfo{person}{Haojun Xia}, \bibinfo{person}{Zhen Zheng}, \bibinfo{person}{Yuchao Li}, \bibinfo{person}{Donglin Zhuang}, \bibinfo{person}{Zhongzhu Zhou}, \bibinfo{person}{Xiafei Qiu}, \bibinfo{person}{Yong Li}, \bibinfo{person}{Wei Lin}, {and} \bibinfo{person}{Shuaiwen~Leon Song}.} \bibinfo{year}{2023}\natexlab{}.
\newblock \bibinfo{title}{Flash-LLM: Enabling Cost-Effective and Highly-Efficient Large Generative Model Inference with Unstructured Sparsity}.
\newblock
\newblock
\showeprint[arxiv]{2309.10285}~[cs.DC]
\urldef\tempurl%
\url{https://arxiv.org/abs/2309.10285}
\showURL{%
\tempurl}


\bibitem[Xu et~al\mbox{.}(2023a)]%
        {xu2023llmcad}
\bibfield{author}{\bibinfo{person}{Daliang Xu}, \bibinfo{person}{Wangsong Yin}, \bibinfo{person}{Xin Jin}, \bibinfo{person}{Ying Zhang}, \bibinfo{person}{Shiyun Wei}, \bibinfo{person}{Mengwei Xu}, {and} \bibinfo{person}{Xuanzhe Liu}.} \bibinfo{year}{2023}\natexlab{a}.
\newblock \bibinfo{title}{LLMCad: Fast and Scalable On-device Large Language Model Inference}.
\newblock
\newblock
\showeprint[arxiv]{2309.04255}~[cs.NI]
\urldef\tempurl%
\url{https://arxiv.org/abs/2309.04255}
\showURL{%
\tempurl}


\bibitem[Xu et~al\mbox{.}(2023b)]%
        {LLMCad}
\bibfield{author}{\bibinfo{person}{Daliang Xu}, \bibinfo{person}{Wangsong Yin}, \bibinfo{person}{Xin Jin}, \bibinfo{person}{Ying Zhang}, \bibinfo{person}{Shiyun Wei}, \bibinfo{person}{Mengwei Xu}, {and} \bibinfo{person}{Xuanzhe Liu}.} \bibinfo{year}{2023}\natexlab{b}.
\newblock \bibinfo{title}{LLMCad: Fast and Scalable On-device Large Language Model Inference}.
\newblock
\newblock
\showeprint[arxiv]{2309.04255}~[cs.NI]
\urldef\tempurl%
\url{https://arxiv.org/abs/2309.04255}
\showURL{%
\tempurl}


\bibitem[Xu et~al\mbox{.}(2024)]%
        {xu2024onebit}
\bibfield{author}{\bibinfo{person}{Yuzhuang Xu}, \bibinfo{person}{Xu Han}, \bibinfo{person}{Zonghan Yang}, \bibinfo{person}{Shuo Wang}, \bibinfo{person}{Qingfu Zhu}, \bibinfo{person}{Zhiyuan Liu}, \bibinfo{person}{Weidong Liu}, {and} \bibinfo{person}{Wanxiang Che}.} \bibinfo{year}{2024}\natexlab{}.
\newblock \bibinfo{title}{OneBit: Towards Extremely Low-bit Large Language Models}.
\newblock
\newblock
\showeprint[arxiv]{2402.11295}~[cs.CL]
\urldef\tempurl%
\url{https://arxiv.org/abs/2402.11295}
\showURL{%
\tempurl}


\bibitem[Xue et~al\mbox{.}(2024)]%
        {xue2024powerinfer2}
\bibfield{author}{\bibinfo{person}{Zhenliang Xue}, \bibinfo{person}{Yixin Song}, \bibinfo{person}{Zeyu Mi}, \bibinfo{person}{Xinrui Zheng}, \bibinfo{person}{Yubin Xia}, {and} \bibinfo{person}{Haibo Chen}.} \bibinfo{year}{2024}\natexlab{}.
\newblock \bibinfo{title}{PowerInfer-2: Fast Large Language Model Inference on a Smartphone}.
\newblock
\newblock
\showeprint[arxiv]{2406.06282}~[cs.LG]
\urldef\tempurl%
\url{https://arxiv.org/abs/2406.06282}
\showURL{%
\tempurl}


\bibitem[Yu et~al\mbox{.}(2022)]%
        {ocra}
\bibfield{author}{\bibinfo{person}{Gyeong-In Yu}, \bibinfo{person}{Joo~Seong Jeong}, \bibinfo{person}{Geon-Woo Kim}, \bibinfo{person}{Soojeong Kim}, {and} \bibinfo{person}{Byung-Gon Chun}.} \bibinfo{year}{2022}\natexlab{}.
\newblock \showarticletitle{Orca: A Distributed Serving System for {Transformer-Based} Generative Models}. In \bibinfo{booktitle}{\emph{16th USENIX Symposium on Operating Systems Design and Implementation (OSDI 22)}}. \bibinfo{publisher}{USENIX Association}, \bibinfo{address}{Carlsbad, CA}, \bibinfo{pages}{521--538}.
\newblock
\showISBNx{978-1-939133-28-1}
\urldef\tempurl%
\url{https://www.usenix.org/conference/osdi22/presentation/yu}
\showURL{%
\tempurl}


\bibitem[Zhang et~al\mbox{.}(2024)]%
        {zhang2024tinyllama}
\bibfield{author}{\bibinfo{person}{Peiyuan Zhang}, \bibinfo{person}{Guangtao Zeng}, \bibinfo{person}{Tianduo Wang}, {and} \bibinfo{person}{Wei Lu}.} \bibinfo{year}{2024}\natexlab{}.
\newblock \bibinfo{title}{TinyLlama: An Open-Source Small Language Model}.
\newblock
\newblock
\showeprint[arxiv]{2401.02385}~[cs.CL]
\urldef\tempurl%
\url{https://arxiv.org/abs/2401.02385}
\showURL{%
\tempurl}


\bibitem[Zhao et~al\mbox{.}(2024)]%
        {MLSYS2024_5edb57c0}
\bibfield{author}{\bibinfo{person}{Yilong Zhao}, \bibinfo{person}{Chien-Yu Lin}, \bibinfo{person}{Kan Zhu}, \bibinfo{person}{Zihao Ye}, \bibinfo{person}{Lequn Chen}, \bibinfo{person}{Size Zheng}, \bibinfo{person}{Luis Ceze}, \bibinfo{person}{Arvind Krishnamurthy}, \bibinfo{person}{Tianqi Chen}, {and} \bibinfo{person}{Baris Kasikci}.} \bibinfo{year}{2024}\natexlab{}.
\newblock \showarticletitle{Atom: Low-Bit Quantization for Efficient and Accurate LLM Serving}. In \bibinfo{booktitle}{\emph{Proceedings of Machine Learning and Systems}}, \bibfield{editor}{\bibinfo{person}{P.~Gibbons}, \bibinfo{person}{G.~Pekhimenko}, {and} \bibinfo{person}{C.~De Sa}} (Eds.), Vol.~\bibinfo{volume}{6}. \bibinfo{pages}{196--209}.
\newblock
\urldef\tempurl%
\url{https://proceedings.mlsys.org/paper_files/paper/2024/file/5edb57c05c81d04beb716ef1d542fe9e-Paper-Conference.pdf}
\showURL{%
\tempurl}


\bibitem[Zhou et~al\mbox{.}(2022)]%
        {Pets}
\bibfield{author}{\bibinfo{person}{Zhe Zhou}, \bibinfo{person}{Xuechao Wei}, \bibinfo{person}{Jiejing Zhang}, {and} \bibinfo{person}{Guangyu Sun}.} \bibinfo{year}{2022}\natexlab{}.
\newblock \showarticletitle{{PetS}: A Unified Framework for {Parameter-Efficient} Transformers Serving}. In \bibinfo{booktitle}{\emph{2022 USENIX Annual Technical Conference (USENIX ATC 22)}}. \bibinfo{publisher}{USENIX Association}, \bibinfo{address}{Carlsbad, CA}, \bibinfo{pages}{489--504}.
\newblock
\showISBNx{978-1-939133-29-11}
\urldef\tempurl%
\url{https://www.usenix.org/conference/atc22/presentation/zhou-zhe}
\showURL{%
\tempurl}


\end{thebibliography}

\appendix

\end{document}